\documentclass{acm_proc_article-sp}

\usepackage{graphicx}
\usepackage{amsmath}
\usepackage{float}
\usepackage{subfigure}
\usepackage{icomma}
\usepackage{url}
\usepackage{xspace}
\usepackage{hyperref}
\usepackage{ifpdf}
\usepackage{xcolor}
\usepackage{cite}

\newcommand{\Posts}{\ensuremath{\mathcal{P}}\xspace}
\newcommand{\PostsSc}{\ensuremath{\mathcal{P}_{\text{science}}\xspace}}
\newcommand{\PostsCon}{\ensuremath{\mathcal{P}_{\text{consp}}\xspace}}
\newcommand{\GlikesP}[1]{\ensuremath{G_{#1}^{\textbf{L}}}}
\newcommand{\VlikesP}[1]{\ensuremath{V_{#1}^{\textbf{L}}}}
\newcommand{\ElikesP}[1]{\ensuremath{E_{#1}^{\textbf{L}}}}
\newcommand{\Glikes}{\ensuremath{G^{\textbf{L}}}}
\newcommand{\Vlikes}{\ensuremath{V^{\textbf{L}}}}
\newcommand{\Elikes}{\ensuremath{E^{\textbf{L}}}}
\newcommand{\GlikesSc}{\GlikesP{\text{science}}}
\newcommand{\VlikesSc}{\VlikesP{\text{science}}}
\newcommand{\ElikesSc}{\ElikesP{\text{science}}}
\newcommand{\GlikesCon}{\GlikesP{\text{consp}}}
\newcommand{\VlikesCon}{\VlikesP{\text{consp}}}
\newcommand{\ElikesCon}{\ElikesP{\text{consp}}}
\newcommand{\NumLikes}[1]{\ensuremath{\theta(#1)}}
\newcommand{\Engagement}[1]{\ensuremath{\psi(#1)}}
\newcommand{\Polarization}[1]{\ensuremath{\rho(#1)}}

\newcommand{\facebook}{Facebook\xspace}

\newcommand{\St}{;\:}
\newcommand{\Degree}[1]{\mathit{deg}(#1)}

\newif\ifshowcomments
\showcommentstrue

\ifshowcomments
\newcommand{\mynote}[2]{\fbox{\bfseries\sffamily\scriptsize{#1}}
 {\small$\blacktriangleright$\textsf{\emph{#2}}$\blacktriangleleft$}}
\else
\newcommand{\mynote}[2]{}
\fi

\begin{document}

\title{Viral Misinformation: The Role of Homophily and Polarization}

\newcommand{\emailFontSize}{11}
\newcommand{\emailBaselineSkip}{13}

\numberofauthors{6} 
%
\author{
\alignauthor
Aris Anagnostopoulos\\
       \affaddr{Dept. of Computer, Control, and Management Engineering}\\
       \affaddr{Sapienza University of Rome}\\
       \affaddr{Italy}
\alignauthor
Alessandro Bessi\\
       \affaddr{IUSS Pavia}\\
       \affaddr{Piazza della Vittoria}\\
       \affaddr{Pavia, Italy}
\alignauthor Guido Caldarelli\\
       \affaddr{IMT Lucca}\\
       \affaddr{Piazza San Ponziano 6}\\
       \affaddr{Lucca, Italy}
\and  
\alignauthor Michela Del Vicario\\
       \affaddr{IMT Lucca}\\
       \affaddr{Piazza San Ponziano 6}\\
       \affaddr{Lucca, Italy}
\alignauthor Fabio Petroni\\
       \affaddr{Dept. of Computer, Control, and Management Engineering}\\
       \affaddr{Sapienza University of Rome}\\
       \affaddr{Italy}\\
%
\alignauthor Antonio Scala\\
       \affaddr{ISC-CNR}\\
       \affaddr{Piazzale Aldo Moro}\\
       \affaddr{Roma, Italy}\\
      	\and
	\alignauthor Fabiana Zollo\\
              \affaddr{IMT Lucca}\\
              \affaddr{Piazza San Ponziano 6}\\
              \affaddr{Lucca, Italy}\\
      \alignauthor Walter Quattrociocchi\titlenote{Corresponding author}\\
       \affaddr{IMT Lucca}\\
       \affaddr{Piazza San Ponziano 6}\\
       \affaddr{Lucca, Italy}\\
       \email{{\fontsize{\emailFontSize}{\emailBaselineSkip}\selectfont walter.quattrociocchi@imtlucca.it}}
      }

\maketitle
\begin{abstract}
The spreading of unsubstantiated rumors on online social networks (OSN)
either unintentionally or intentionally (e.g., for political reasons or
even trolling) can
have serious consequences such as in the recent case of rumors about Ebola
causing disruption to health-care workers.
Here we show that indicators aimed at quantifying information
consumption patterns might provide important insights about the
virality of false claims.
In particular, we address the driving forces behind the popularity of contents by
analyzing a sample of $1.2M$ Facebook Italian users consuming
different (and opposite) types of information (science and conspiracy
news). We show that users' engagement across different
contents correlates with the number of friends having similar consumption
patterns ({\em homophily}), indicating the area in the social network
where certain types of contents are more likely to spread. Then, we test
diffusion patterns on an external sample of $4,709$ intentional
satirical false claims showing that neither the presence of hubs
(structural properties) nor the most active users (influencers) are
prevalent in viral phenomena. Instead, we found out that in an
environment where misinformation is
pervasive, users' aggregation around shared beliefs may make the usual
exposure to conspiracy stories ({\em polarization}) a determinant for
the virality of false information.
\end{abstract}


\terms{Misinformation, Virality, Attention Patterns}


\section{Introduction}
The Web has become pervasive and digital technology permeates every
aspect of daily life.
Social interaction, healthcare activity, political engagement, and
economic decision-making are influenced by the digital hyperconnectivity
\cite{Lotan11,Lewis2012,Leskovec2010,kleinberg2013analysis,Richard2004,howard11-1,Moreno2011,Bond2012}.
Nowadays, everyone can produce and access a variety of information actively participating in the diffusion and reinforcement of
narratives, by which we mean information coherent with a given worldview.
Such a shift of paradigm in the consumption of information has profoundly affected the way people get informed\cite{Buckingham2012,Centola2010,eRep,QuattrociocchiCL11,QuattrociocchiPC09,Bekkers2011,Quattrociocchi2014}.
However, the role of the socio-technical systems in shaping the public
opinion still remains unclear. Indeed, the spreading of unsubstantiated rumors, 
whether it is unintentional or intentional (e.g., for political reasons or
even trolling), could have
serious consequences; the World Economic
Forum has listed \emph{massive digital
misinformation} as one of the main risks for the modern society
\cite{Davos13}.

On online social networks, users discover and share information with
their friends and through {\em cascades} of reshares information might
reach a large number of individuals.
Cascades are rare \cite{Goel2012}, but are common on online social media
such as Facebook or Twitter \cite{DAF13,Kumar2010,MaSC13}.
Interesting is the popular case of Senator Cirenga's
\cite{Cirenga1,Cirenga2} law proposing to fund policy makers with 134
million of euros (10\% of the Italian GDP) in case of defeat in the
political competition.
This was an intentional joke -- the text of the post was explicitly
mentioning its provocative nature -- that became popular within online
political activists to an extent that it has been used as an
argumentation in political debates \cite{forconi_cirenga2013}.
The information overload mixed with users' limited attention may facilitate the creation of urban legends.

Predicting the evolution of cascades is far from trivial
\cite{Watts2006}. In this work, we want to understand the driving forces behind the
popularity of contents accounting for the users information consumption
patterns.
To this end, we focus on two very distinct types of information:
scientific and conspiracy news.
The former are aiming at diffusing scientific knowledge as well as
scientific thinking, while the latter provide alternative arguments that
are difficult to substantiate.
Conspiracists tend to reduce the complexity of reality by explaining
significant social or political occurrences as plots conceived by
powerful individuals or organizations. Since these kinds of arguments can
sometimes involve the rejection of science, alternative explanations are
invoked to replace the scientific evidence \cite{Bauer2013}.
For instance, people who reject the link between HIV and AIDS generally
believe that AIDS was created by the U.S. Government to control the
African American population \cite{Sunstein12,mckelvey2013truthy}.
Just recently, the fear of an Ebola outbreak in the United States rippled
through social media networks
\cite{RumorEbola1,RumorEbola2,RumorEbola3}. 
Furthermore, the spread of misinformation
can be particularly difficult to correct \cite{Meade2002,Mann2000,Garrett2013,SOCINFO14}. 
In fact, recently, in \cite{Mocanu2014} it has been shown that conspiracist and
mainstream information reverberate in a similar way on social media and
that users generally exposed to conspiracy stories are the more prone to
like and share satirical information \cite{Bessi2014}.
Meanwhile, scientific debunking on conspiracy rumors creates a
reinforcement effect in consuming conspiracist information on polarized
users \cite{SOCINFO14}.

\textbf{Virality of unsubstantiated rumors.} In this work, we analyze a
sample of $1.2M$ Facebook Italian users consuming different (and
opposite) kind of information: scientific and conspiracy news.
We observe that users engagement across different contents
correlates with the number of friends having similar consumption patterns
(\emph{homophily}). Then, we focus on the popularity of posts of the
two categories, finding that both present similar statistical
signatures.
Finally, we test this diffusion patterns on an external sample of $4,709$
intentional false information -- satirical version and paradoxical
claims -- confirming that neither hubs (structural properties) nor the
most active users (influencers) are relevant to characterize viral phenomena.
Our analysis shows that where misinformation is pervasive, 
users' aggregation around shared beliefs makes the frequent
exposure to conspiracy stories ({\em polarization}) a determinant for
the virality of false information.
Furthermore, we provide important insights toward the
understanding of viral processes around false information on online
social media. In particular, through the relation between the users'
engagement and the number of friends, we provide new metrics -- i.e.,
users polarization defined on the information consumption patterns -- to
detect areas in the social network where false claims are more likely to
spread.
We conclude the paper showing a nice case study about the role of
the frequent exposure to conspiracy stories (\emph{polarization}) on
the virality of false information.

\section{Related Works}

Recent studies explored the evolution of social phenomena on online social networks from observable data; one of the most investigated aspects are the structural properties and their effect on social dynamics~\cite{Centola2010,ugander2012,Cen3}. 
The strength of weak ties is currently matter of debate. In fact, in \cite{Cen3} it is shown that while long ties are relevant  in the spreading of innovation and social movements, they are not enough to spread social reinforcement.
In~\cite{Centola2010} the author found evidence of an effect of reinforcement in the adoption for clustered networks, where many redundant ties are present. 
In a more recent work~\cite{ugander2012}, the number of different connected components, representing separate social contexts for the user, proved to be more influential than the number of friends itself and other properties of potential influencers have been addressed in~\cite{Aral2}.
Another interesting challenge is the one studying the possibility to distinguish influence from homophily \cite{Aral,Ada1,Cen2}.
In ~\cite{Aral} it is performed over the global network of instant messaging traffic among about 30 millions users on Yahoo.com, with complete data on the adoption of a mobile service application and precise dynamic behavioral data on users. The effect of homophily proves to explain more than $50\%$ of the phenomena perceived as contagion. 
In~\cite{Ada1} the role of social network and exposure to friends' activities in information re-sharing on Facebook is analyzed. Through controlled experiments in which user were divided in exposed and not exposed to friends' re-sharing, authors were able to isolate contagion from other confounding effects like homophily. They claimed that in the exposed case there is a considerably higher chance to share contents. 

Recent studies settled on Facebook aimed at unfolding cascades characteristics~\cite{DAF13} and predicting their trajectories and shapes~\cite{Ada2}. As for the characteristics, it is found that a small but significant fraction of posts forms wide and deep cascades and that different cascades may evolve in different ways. 
Many aspects of cascades' behavior -- e.g., under which structural and user-constrained properties is  possible to predict them -- are hard tasks that have not been completely exploited.
In the last years, a new online phenomenon is attracting the interest of the researchers community, the spreading of unsubstantiated and false claims through OSN (as Facebook), that often reverberate leading to mass misinformation.
The study in~\cite{Mocanu2014} is a detailed analysis of the information consumption by Facebook users on different categories of pages: alternative information sources, political activism and mainstream media. Authors pointed out evidences that mainstream media information reverberate as long as unsubstantiated one, and that the exposition to the latter makes users more likely to interact with intentionally injected false information. More recently, in \cite{SOCINFO14} it has been shown that the exposure to debunking posts might increase the engagement of users in consuming conspiracy information.

\section{Data Collection}
\label{sec:data}

In this work, we aim at testing the relationship of content consumption
with respect to social networks' structure and viral phenomena. 
To do this, we need to clearly identify sources that are promoting different contents 
referring to opposite worldviews. 
We define the space of our investigation with the help of
\facebook groups very active in the debunking of conspiracy theses (see
the acknowledgements section for further details).
We started from 73 public \facebook pages, from which 34 are about
scientific news and 39 about news that can be considered conspiratorial;
we refer to the former as \emph{science pages} and to the latter as
\emph{conspiracy pages}.
In Table~\ref{tab:data_dim} we summarize the details of our data
collection. We downloaded all
posts from these pages in a timespan of 4 years (2010 to 2014). In
addition, we collected all the \emph{likes} and \emph{comments} from the
posts, and we counted the number of \emph{shares}\footnote{Unfortunately,
although \facebook provides this number for a given post, it provides
the IDs of only a small number of the sharers.}. In total, we collected around $9M$ likes and $1M$
comments, performed by about $1.2M$ and $280K$ \facebook users, respectively (see
Table~\ref{tab:data_dim}). Likes, shares, and comments have a different
meaning from the user viewpoint.  Most of the times, a
like stands for a positive feedback to the
post; a share expresses the will to increase the visibility of a
given information; and a comment is the way in which online
collective debates take form. Comments may contain negative or positive
feedbacks with respect to the post.

In addition, we collected the ego networks of users who liked at least one
post on science or conspiracy pages\footnote{We used publicly available
data, so we collected only
data for which the users had the corresponding permissions open. We
estimated that these form a percentage of more than $96\%$ of the total
connections, allowing us to perform a meaningful analysis regarding the
connections between users.}, that
is, for each of these users we collected the list of friends and the links between
them. This allowed us to create a social network of users and the
(publicly declared) connections between them for a total of about $1.2M$
nodes and $34.5M$ edges.
%
\begin{center}
\begin{table}[!h]
\centering
\small
\begin{tabular}{l|c|c|c}
\hline\bf {  }  & \bf {Total} & \bf {Science} & \bf {Conspiracy}  \\ \hline 
Pages & $ 73 $ & $ 34 $ & $ 39 $ \\
Posts & $ 271,296 $ & $ 62,705 $ & $ 208,591 $ \\
Likes & $ 9,164,781 $ & $ 2,505,399 $ & $ 6,659,382$  \\
Comments & $1,017,509 $ & $ 180,918 $ & $ 836,591 $\\
Shares &  $17,797,819$ & $1,471,088$ & $16,326,731$\\
Likers & $ 1,196,404 $ & $ 332,357 $ & $ 864,047 $\\
Commenters & $ 279,972 $ & $ 53,438 $ & $ 226,534 $\\
\end{tabular}\newline
  \caption{ \textbf{Breakdown of Facebook dataset.} The number of pages,
    posts, likes, comments, and shares for science and conspiracy pages.}
  \label{tab:data_dim}
\end{table}
\end{center}

Finally, we obtained access to $4,709$ posts from two satirical
\facebook pages (to which we will refer to as \emph{troll posts} and
\emph{troll pages}) promoting intentionally
false and caricatural version of the most debated issues.
The more popular of the two is called called ``Semplicemente Me'', it 
is followed by about $7K$ users and it is focused on general online
rumors. The second one is called ``Simply Humans'', it is followed by
about $1K$ users, and it hosts mostly posts of conspiratorial nature.
We collected about $40K$ likes and $59K$ comments on these posts,
performed by about $16K$ and $43K$ \facebook users, respectively.
These pages were able to trigger several viral phenomena, with one of
them reaching more than $100K$ shares. In Section~\ref{sec:virality},
we use troll memes to measure how the social ecosystem under
investigation is responding to the injection of false information.

\section{Preliminaries and Definitions}
\label{sec:prelim}

Here we describe how we preprocess our dataset and provide some of the basic
definitions that we use throughout the entire paper.
Let $\Posts$ be the set of all the posts in our collection, and define
$\PostsSc$ ($\PostsCon$) as the set of posts posted on one of the 34 (39)
\facebook pages about science (conspiracy) news. 
We call posts that were posted in pages about science, \emph{science posts}, and similarly for
\emph{conspiracy posts}.
Let $V$ be the set of all the $1.2M$ users that we observed and $E$ the
set of edges representing the \facebook friendship connections between
them. These define a graph $G=(V,E)$.

We also define the graph of likes, $\Glikes=(\Vlikes,\Elikes)$, which is
the subgraph of $G$ composed of users who have liked at least one post.
Thus, $\Vlikes$ is the set of users of $V$ who have liked at least one
post, and we set $\Elikes=\{(u,v)\in E\St u,v\in\Vlikes\}$.

For each user $u$ let
$\NumLikes{u}$ be the total number of likes that $u$
has expressed in posts in our collection~$\Posts$. We define
$\Engagement{u}$ and call it the 
\emph{engagement} of user $u$ as the normalized liking activity with
respect to of all the users in our dataset, namely
$\Engagement{u}=\frac{\NumLikes{u}}{\max_v\NumLikes{v}}$.
We are interested in learning the extent to which users are polarized in a given content.
Following previous works~\cite{Mocanu2014,Bessi2014,SOCINFO14}, 
we are interested in studying the polarization of users, which we informally define as the tendency of
users to interact about only with a single type of information. Here we study
the polarization towards science and conspiracy. To quantify it, we define 
the \emph{polarization} of user $u\in\Vlikes$, $\rho(u)\in [0,1]$, as the
ratio of likes that $u$ has performed on conspiracy posts: assuming that $u$ has performed $x$ and $y$ likes on science and
conspiracy posts, respectively, we let $\rho(u)=y/(x+y)$. Thus, a user
$u$ for whom $\rho(u)=0$ is polarized towards science, whereas a user
with $\rho(u)=1$ is polarized towards conspiracy.
Note that we use the liking activity to define polarization and we
ignore the commenting activity. The former is usually an explicit
endorsement of the original post, whereas a comment may be an
endorsement, a criticism, or even a response to a previous comment.

\begin{figure}
 	\centering
       \includegraphics[width=0.5\textwidth]{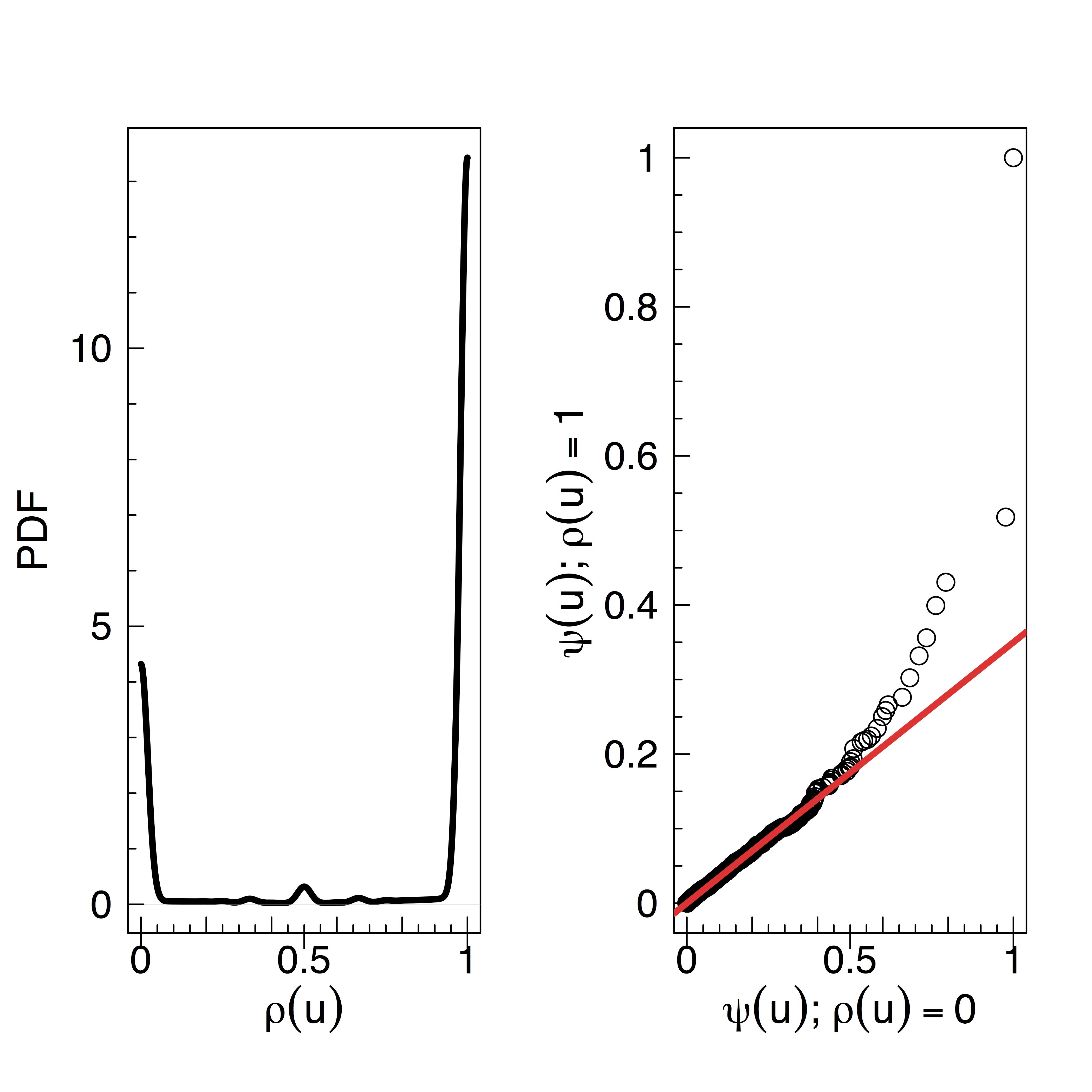}
	   \caption{\textbf{Polarization on contents}. Probability
	     density function (PDF) of users' polarization across posts
	       and quantile--quantile comparison (QQ plot) between the
	       distributions of the values
$\{\Engagement{u}\St \Polarization{u}=0\}$ and
$\{\Engagement{u}\St \Polarization{u}=1\}$.}
\label{fig:polarization_dist}
\end{figure}

In Figure~\ref{fig:polarization_dist} we depict the polarization of all
the users in $\Vlikes$. On the left panel, we show the probability density
function (PDF). Recall that value $\rho(u) = 0$
corresponds to users who have liked only science posts and value  $\rho(u) = 1$ to users
who have liked only conspiracy posts. From the plots it is evident that the
vast majority of the users are polarized either towards science or
towards conspiracy. 
Moreover, the
quantile--quantile comparison shows that distributions of the
number of likes for users totally polarized towards science and
conspiracy -- i.e, the distributions of the values
$\{\Engagement{u}\St \Polarization{u}=0\}$ and
$\{\Engagement{u}\St \Polarization{u}=1\}$ -- are
very similar.
To further details on the interaction between users on the two
categories refer to~\cite{Bessi2014}. Findings
depicted in Figure~\ref{fig:polarization_dist} suggest that most of the
likers can be divided into two sets, those \emph{polarized towards
science} and
those \emph{polarized towards conspiracy} news. 
Let $\VlikesSc$ be the users with polarization more than $95\%$ 
towards science (i.e., less than $5\%$
towards conspiracy), formally:
\[\VlikesSc=\{u\in\Vlikes\St \Polarization{u}<0.05\},\]
and $\VlikesSc$ the users with polarization more than $95\%$ towards
conspiracy, namely 
\[\VlikesCon=\{u\in\Vlikes\St \Polarization{u}>0.95\}.\]
For convenience, we call the first set of users \emph{science users} and
the second one \emph{conspiracy users}, without implying that the former
are scientists or that the latter necessarily believe in conspiracy
theories.
Note that these two sets contain most of the users in the peaks in
Figure~\ref{fig:polarization_dist}. We also experimented with values
ranging from $50\%$ to $99\%$, and the results were qualitatively the
same.
We can then define the induced subgraphs of $G$,
$\GlikesSc=(\VlikesSc,\ElikesSc)$ and
$\GlikesCon=(\VlikesCon,\ElikesCon)$ in the natural way, for example,
$\ElikesSc$ contains all the edges in $E$ such that both endpoints are
in $\VlikesSc$.
In Section~\ref{sec:homophily} we study those two graphs extensively.

We also define some of the graph-theoretic terms that we use throughout
the paper. In general, nodes in the graphs represent the
\facebook users of our study. For convenience, we use the terms
\emph{node} and \emph{user} interchangeably.
We assume that a graph to which we refer is clear from the context.
The \emph{degree} of node $u$, $\Degree{u}$, is the number
of neighbors of node $u$. The \emph{clustering coefficient} of $u$ is
the number of closed triplets over the total number of triplets (we use
the time-efficiently method).
The \emph{$k$-core} of a graph $H$
is the maximal subgraph $H'$ of $H$ such that the degree of each node in $H'$ is at
least~$k$. The \emph{coreness} of a vertex is $k$ if it belongs to the $k$-core
but not to the $(k+1)$-core.
The \emph{density} of a graph with $x$ nodes and $y$ edges is
$y/\binom x2$: the ratio of the number of existing edges over
all the possible edges.

Finally, we introduce some statistical notions that we use in
Section~\ref{sec:user-contents} to analyze how science and
conspiracy information is consumed.
Let us define a random variable $T$ on the interval $[0,\infty)$,
indicaticating the time an event takes place (we will use it to
represent the time elapsed from the time a post was posted till a given
user liked it). The cumulative
distribution function (CDF), $F(t) = \textbf{Pr}(T \leq t)$,
indicates the probability that the event takes place within a given time~$t$. The
\emph{survival function}, defined as the complementary CDF (CCDF)\footnote{We remind
that the CCDF of a random variable $X$ is one minus the CDF, 
the function $f(x)=\textbf{Pr}(X>x)$.} of $T$,
gives the probability that an event lasts beyond a given time period $t$.
To estimate this probability we use the \emph{Kaplan--Meier estimate}~\cite{KM58}. 
Let $n_{t}$ denote the number of posts being liked before a given time
step $t$, and let $d_{t}$ denote the number of posts stop being liked
at~$t$. Then, the estimated survival probability after time $t$ is
defined as $(n_{t} - d_{t})/n_{t}$. 
Then, if we have $N$ observations at times $t_1\le t_2\le\dots\le t_N$,
assuming that the events at times $t_i$ are jointly independent, the Kaplan--Meier
estimate of the survival function at time $t$ is defined as
$\hat{S}(t) = \prod_{t_{i}<t}( \frac{n_{t_i} - d_{t_i}}{n_{t_i}})$.

\section{Users and Contents}
\label{sec:user-contents}

\begin{figure}[htbp]
\begin{center}
\includegraphics[width = .5\textwidth]{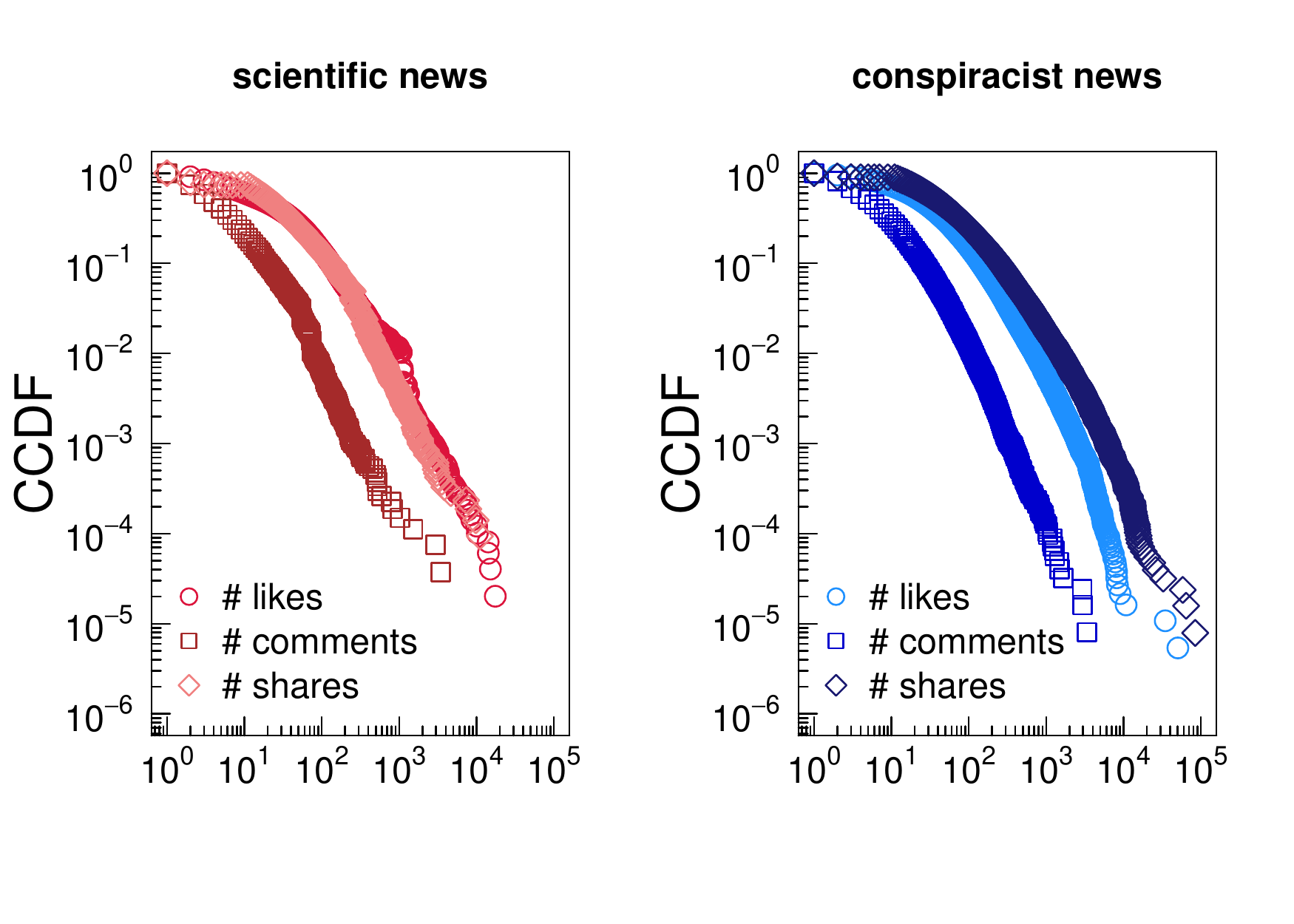}
\caption{Consumption Patterns. Empirical Complementary Cumulative Distribution Function (CCDF) for the number of likes, comments, and shares of $\PostsSc$ and $\PostsCon$.}
\label{fig:consumption}
\end{center}
\end{figure}

\begin{figure}[htbp]
\begin{center}
\includegraphics[width = .4\textwidth]{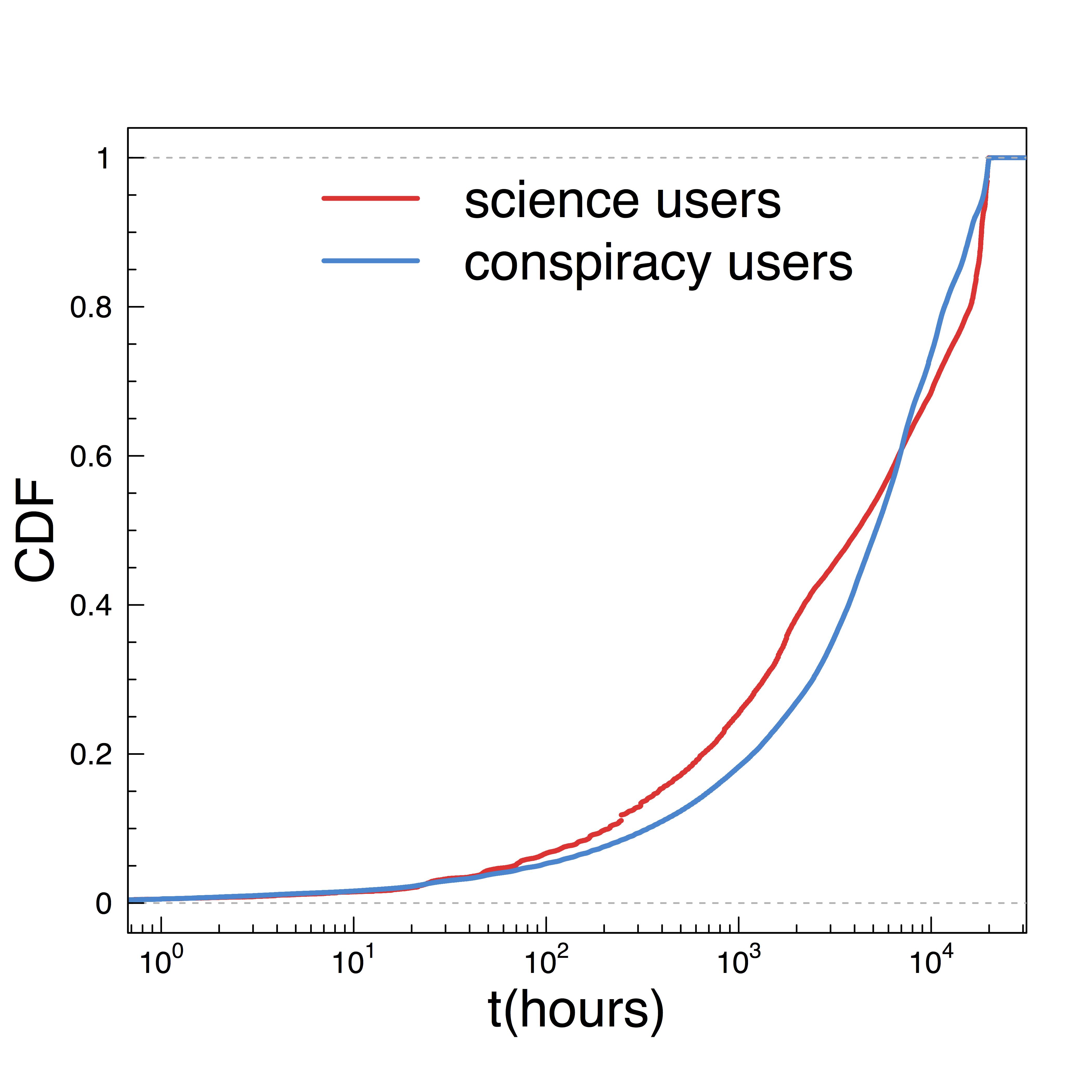}
\caption{\textbf{Users' Lifetime.} Empirical Cumulative Distribution
Function (CDF) for the temporal distance between the first and 
last like of users in $\VlikesSc$ and $\VlikesCon$ on posts in
$\PostsSc$ and $\PostsCon$, respectively; see text for more details.}
\label{fig:users_lifetime}
\end{center}
\end{figure}

\begin{figure}[htbp]
\begin{center}
\includegraphics[width = .4\textwidth]{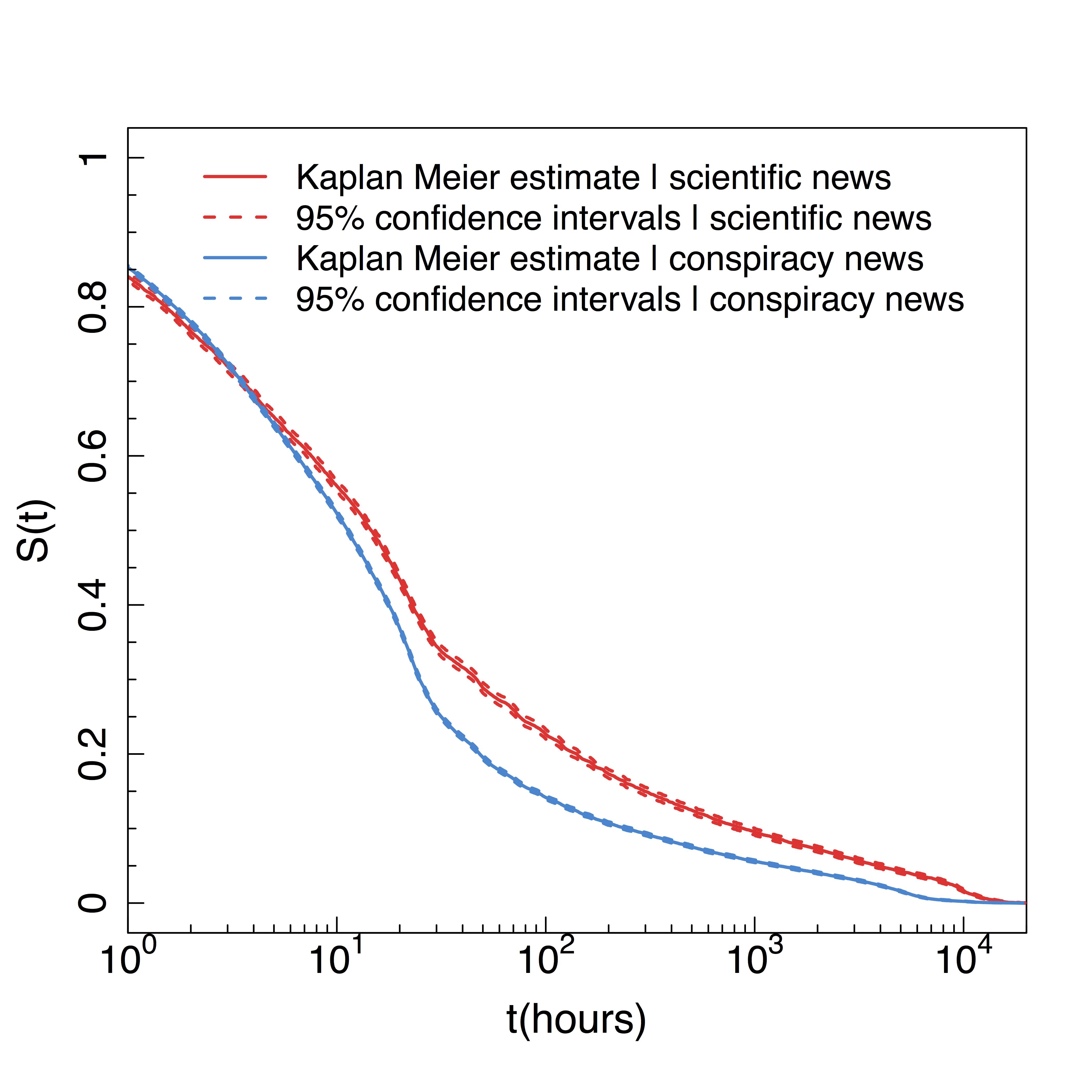}
\caption{\textbf{Posts' Lifetime.} Kaplan--Meier estimate and 95\%
confidence intervals for the Survival Function of $\PostsSc$ and
$\PostsCon$.}
\label{fig:kaplan_meier}
\end{center}
\end{figure}

In this section, we study how users consume information from science and
conspiracy pages.
Figure~\ref{fig:consumption} shows the empirical complementary
cumulative distribution function (CCDF) of the number of likes,
comments, and shares of posts in
$\PostsSc$ and $\PostsCon$.
Now, we concentrate on the users and we measure how the science and
conspiracy users (i.e., users in $\VlikesSc$ and $\VlikesCon$)
keep on consuming information in $\PostsSc$ and $\PostsCon$ over time.
Figure~\ref{fig:users_lifetime} shows the empirical cumulative
distribution function (CDF) for the lifetime of $\VlikesSc$ and
$\VlikesCon$ on posts in $\PostsSc$ and $\PostsCon$, respectively.
The \emph{lifetime} of a user $u\in\VlikesSc$ is the difference of the
post time of the last and first post in $\PostsSc$ that she liked,
and similarly for the lifetime of users in $\VlikesCon$.
Finally, we put our attention on the lifetime of $\PostsSc$ and $\PostsCon$ by
computing their empirical Survival Function.
Figure~\ref{fig:kaplan_meier} shows the Kaplan--Meier estimate and 95\% confidence
intervals for the Survival Function of $\PostsSc$ and $\PostsCon$.
Despite the very different kind of content, the results indicate that
consumption patterns (number of likes, comments, and shares),
as well as persistence of users and posts over time, both categories have similar consumption patterns.

\section{Contents and friends}
\label{sec:homophily}

\begin{figure} 
\centering
\includegraphics[width=0.5\textwidth]{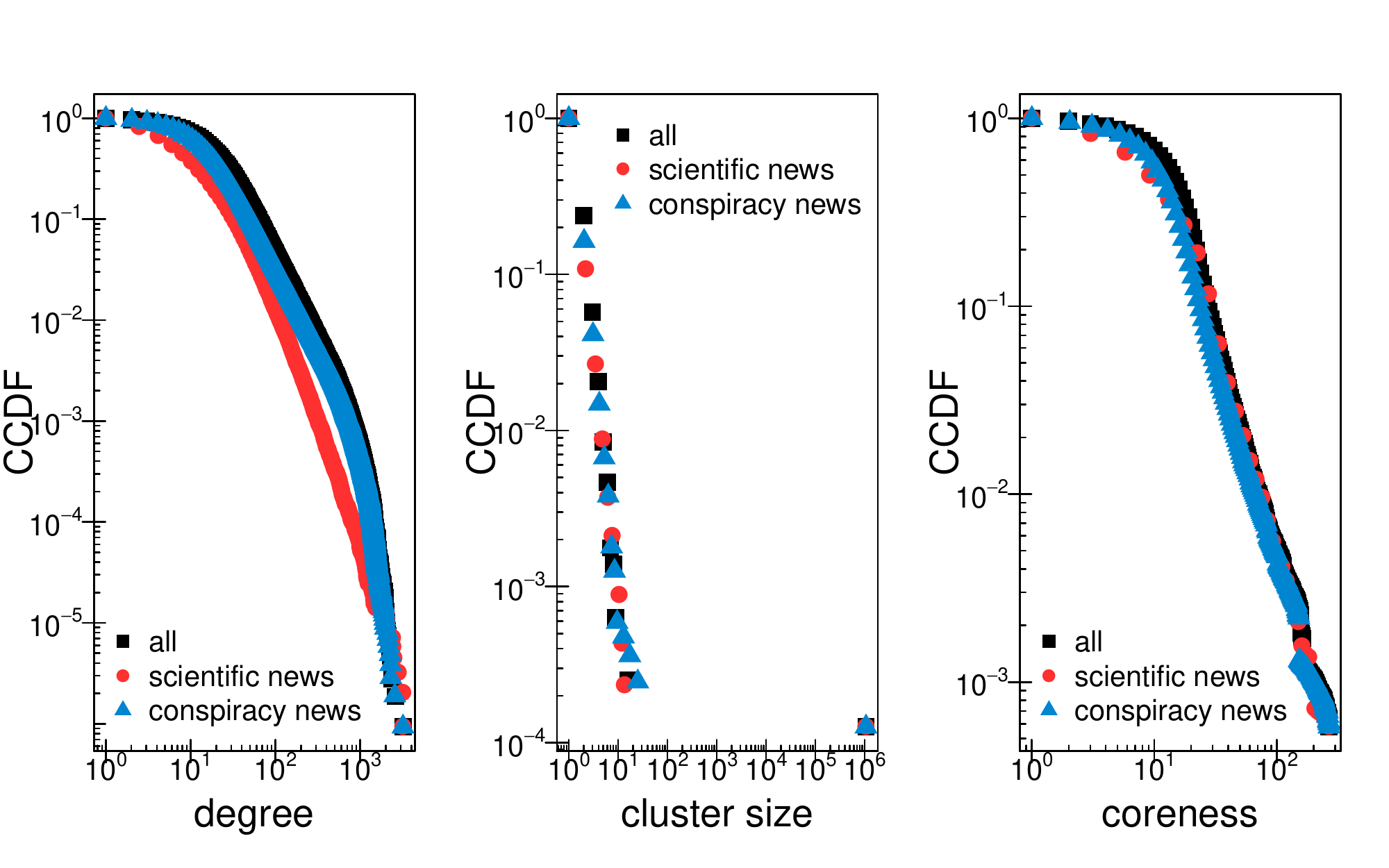}
\caption{Network metrics. Complementary Cumulative Distribution
Function (CCDF) of the degree of each node, the size of the connected
components, and the coreness of each node of the graphs $G$ (all),
$\GlikesSc$ (scientific news), and $\GlikesCon$ (conspiracy news). All graphs present similar distributions.}
\label{fig:basic_metrics}
\end{figure}

\begin{figure} 
\centering
\includegraphics[width=0.5\textwidth]{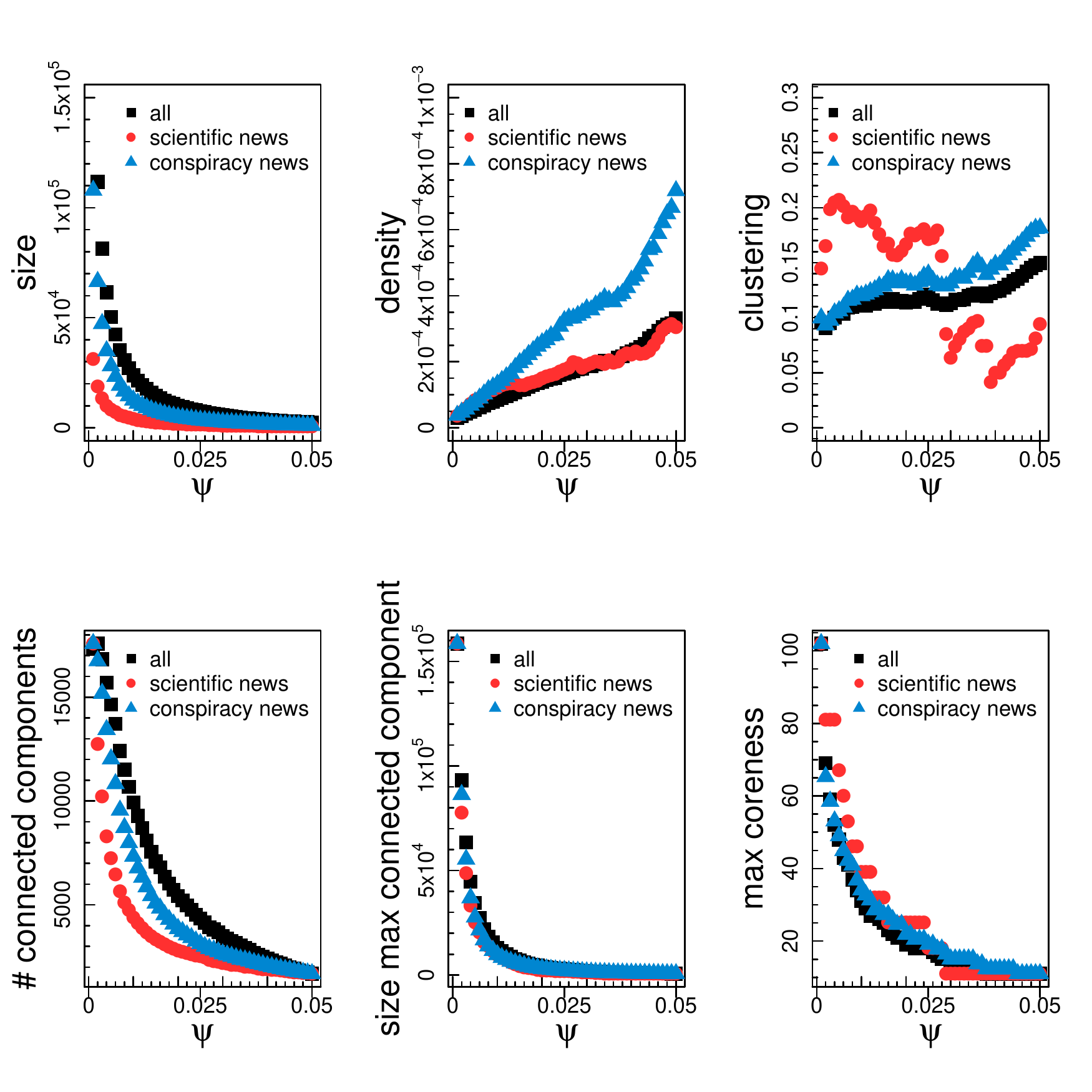} 
\caption{Network topology and users engagement. Size (number of nodes),
density, clustering coefficient, number of connected components, size of
the maximum connected component, and maximum coreness over all the nodes in
the subgraph induced by nodes of engagement $\Engagement{u}\ge x$.}
\label{fig:metrics1}
\end{figure}

In this section, we analyze the topological features of the social
networks by accounting for users' information consumption patterns.
As a first measure, in Figure~\ref{fig:basic_metrics}, we show the
complementary cumulative distribution function (CCDF) of the degree of each node, 
the size of the connected components, and the coreness of each node for the graphs $G$, $\GlikesSc$, and
$\GlikesCon$.  All graphs present similar distributions.
Next we compare the network structure of the entire graph
$G$, and the subgraphs of the polarized users, $\GlikesSc$
and~$\GlikesCon$, when we restrict to different levels of engagement.
In Figure~\ref{fig:metrics1} we show various topological metrics for
subgraphs of the three aforementioned graphs. 
In the $x$-axis we vary the engagement of nodes and consider the subgraphs of $G$,
$\GlikesSc$, and $\GlikesCon$ induced by nodes $u$ with engagement $\Engagement{u}\ge x$. 
In the $y$-axis we consider a variety of topological measures: size (number of nodes),
density, clustering coefficient, number of connected components, size of
the maximum connected component, and maximum coreness over all the nodes in
the subgraph.
The results show that the more we consider active users the more the
connectivity patterns move towards a denser and clustered network.

\begin{figure}
 \centering
       \includegraphics[width=0.4\textwidth]{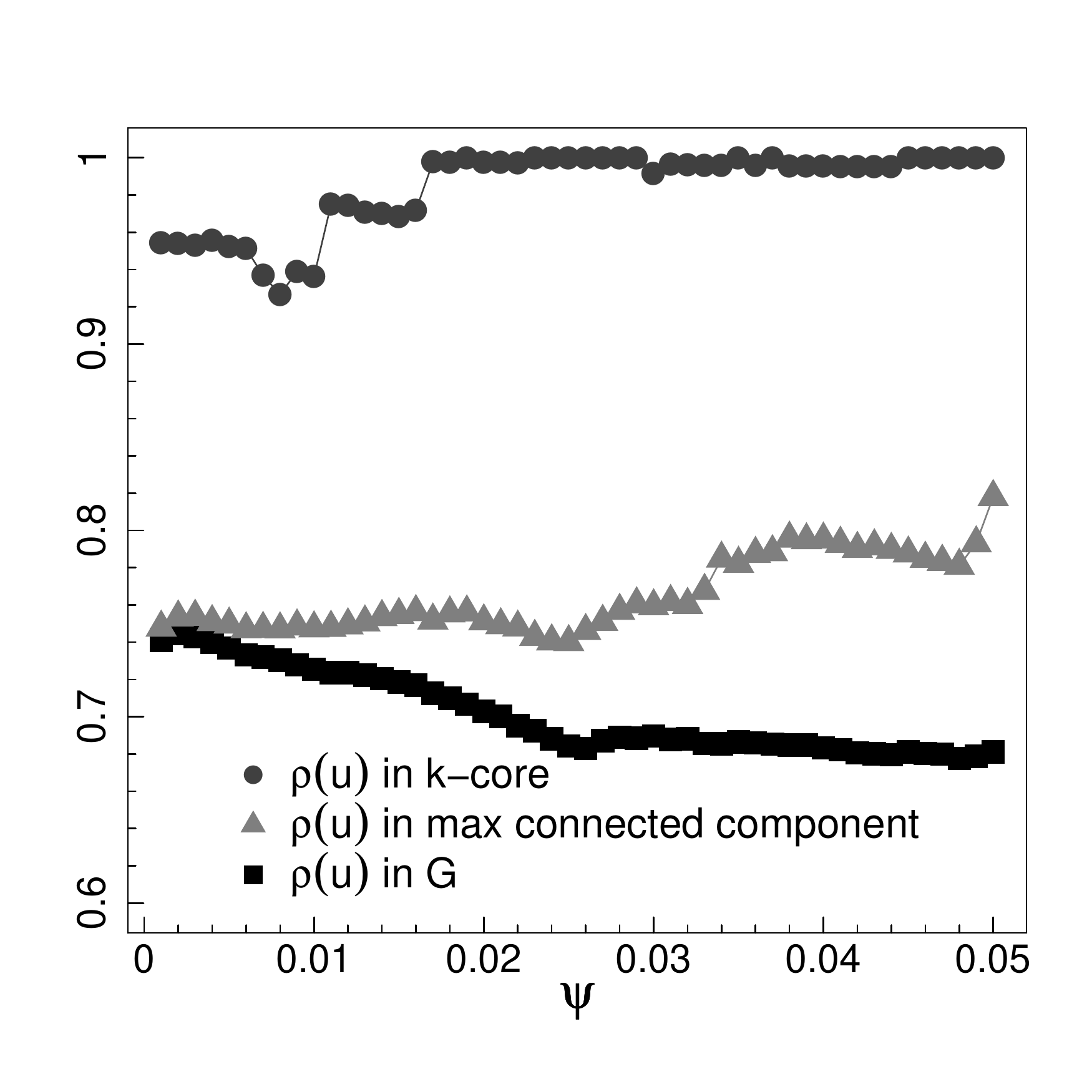}
\caption{Users Engagement and Network. For each value of $x$, we
consider the subnetwork of $G$ induced by users $u$ with
$\Engagement{u}\ge x$ and we compute the average polarization among the
users $u$ in the entire subnetwork, in the maximal $k$-core, and in the
maximal connected component.}
\label{fig:polarization_network_metrics}
\end{figure}

The liking activity is a good approximation to associate users with
their preferred contents~\cite{Mocanu2014,Bessi2014,SOCINFO14},
so users' liking activity across contents of the different categories may be intended as the
preferential attitude towards the one or the other type of information. 
To see how consumers of different contents are distributed over the network,
we measure how the polarization of the network changes along with the users' engagement.
In Figure~\ref{fig:polarization_network_metrics}, we show the average
value of $\Polarization{u}$ as a function of their $\Engagement{u}$:
for each value of $x$, we
consider the subnetwork of $G$ induced by users $u$ with
$\Engagement{u}\ge x$ and we compute the average polarization among the
users $u$ in the entire subnetwork, in the maximal $k$-core, and in the
maximal connected component.

\begin{figure} 
 \centering
\includegraphics[width=0.5\textwidth]{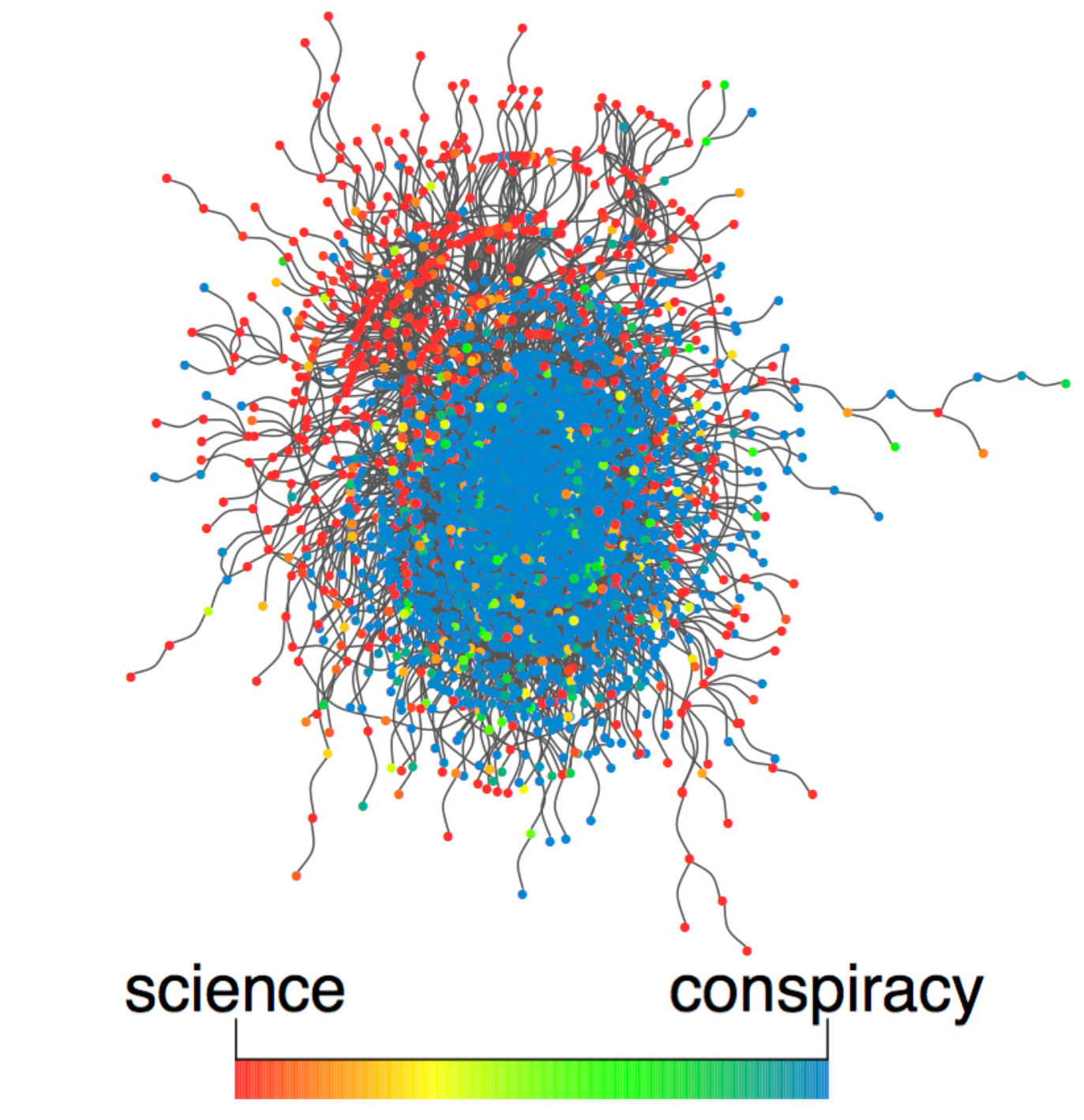}
\caption{Largest connected component. Maximal connected
component of the friendship network for nodes $u$ with $\Engagement{u} > 0.025$, where nodes
are colored according to the polarization $\Engagement{u}$.}
\label{fig:network_plot}
\end{figure}

We can observe in Figure~\ref{fig:polarization_network_metrics} that the
lowest value of the polarization in the largest connected components is
attained at $\Engagement{u} = 0.025$. Given this observation,
in Figure~\ref{fig:network_plot}, we show the largest connected
component of the subnetwork induced by the friendship network $G$ if we
only consider nodes $u\in G$ with $\Engagement{u}>0.025$, where the
color of each node $u$ depends on $\Polarization{u}$ (red nodes are
science users and blue are conspiracy ones).
We identify a main community of polarized users in conspiracy news. 

\begin{figure} 
 \centering
       \includegraphics[width=0.5\textwidth]{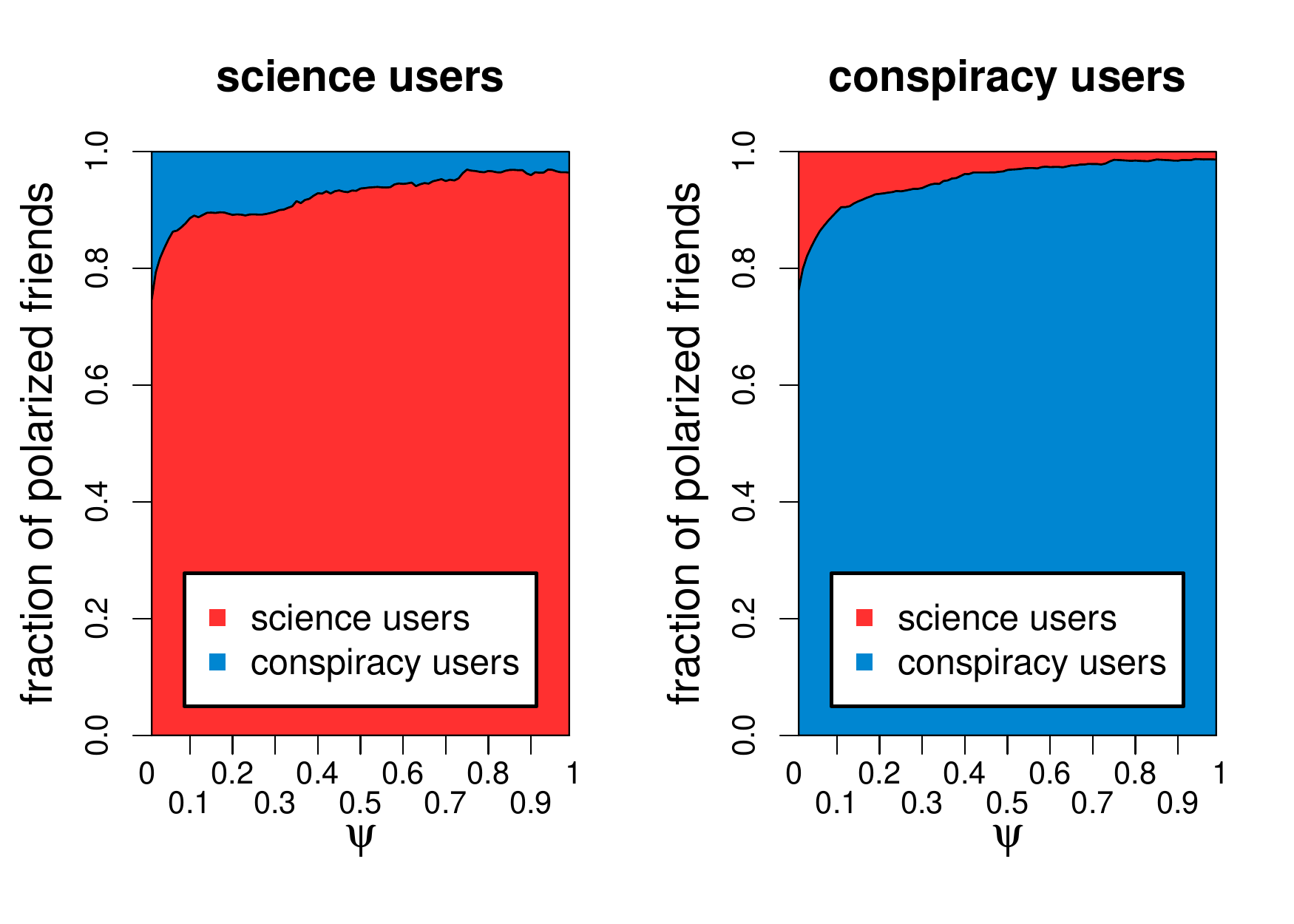}
\caption{Fraction of polarized neighbors as a function of $\Engagement{\cdot}$.} 
\label{fig:homophily}
\end{figure}

\begin{figure}
 \centering
       \includegraphics[width=0.5\textwidth]{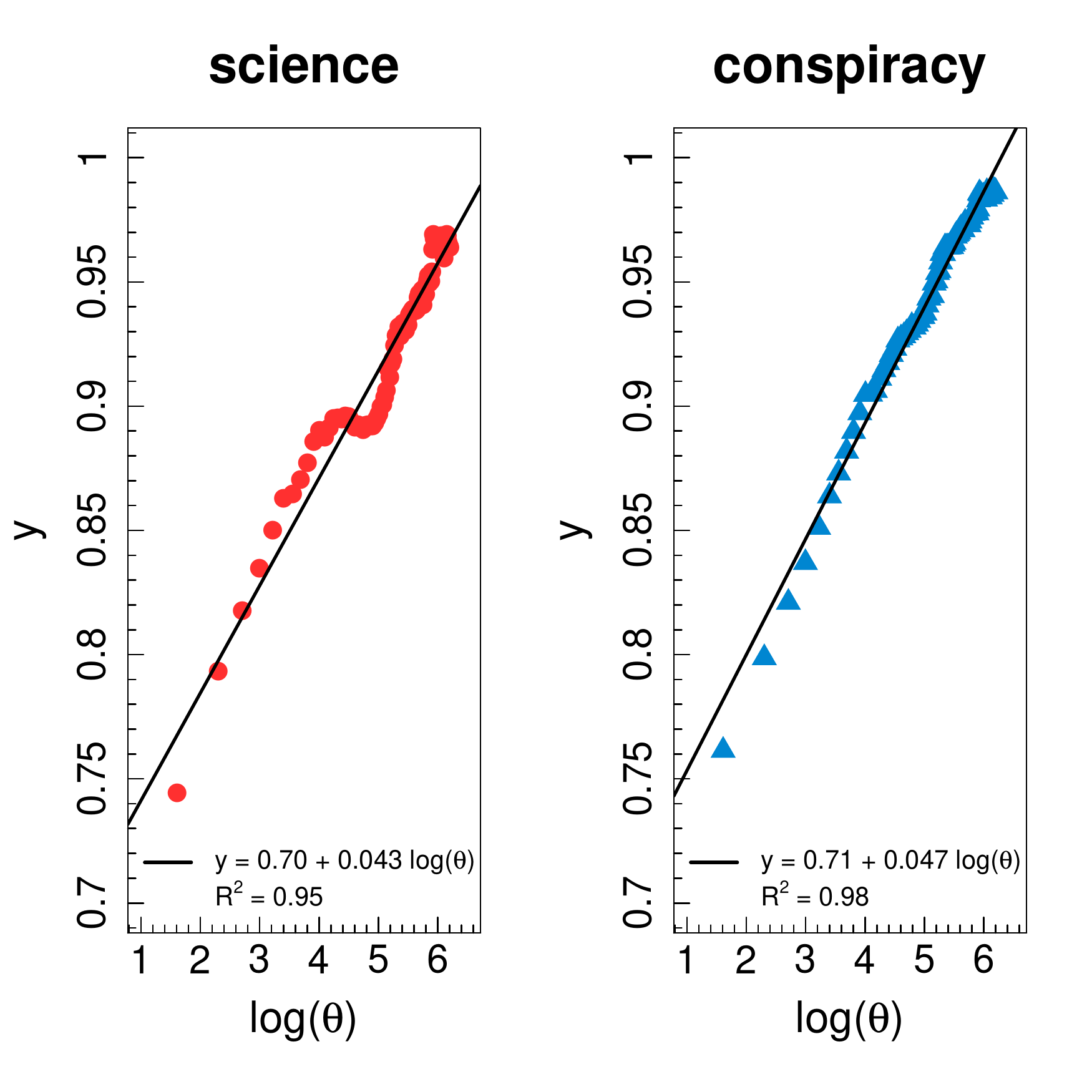}
\caption{Predicting the number of polarized friends. For a polarized user $u$, the fraction of polarized friends in her category $y(u)$ 
can be predicted by means of a linear regression model with intercept
where the explanatory variable is a logarithmic transformation of the
number of likes $\NumLikes{u}$, that is, $y(u) = \beta_{0} +
\beta_{1}\log(\NumLikes{u}).$
Coefficients are estimated using ordinary
least squares and they are---with the corresponding standard
errors inside the round brackets---$\hat{\beta}_{0} = 0.70$ $(0.005)$ and
$\hat{\beta}_{1} = 0.043 $ $(0.001)$, with $R^{2}=0.95$, for users
polarized towards science, and $\hat{\beta}_{0} = 0.71$ $ (0.003)$ and
$\hat{\beta}_{1} = 0.047$ $ (0.0006)$, with $R^{2}=0.98$, for users
polarized towards conspiracy. 
All the p-values are close to zero.}
\label{fig:lr_homophily}
\end{figure}

Figure~\ref{fig:network_plot} hints that there is a strong presence
of homophily with respect to polarization, and we delve more into this
issue. 
In Figure~\ref{fig:homophily} we show the fraction of polarized
friends as a function of the engagement degree~$\Engagement{\cdot}$.
Specifically, in the left panel, we consider the
set of each user $u\in\VlikesSc$ with a given $\Engagement{u}$, we compute the
fraction of $u$'s polarized neighbors in $G$ that are polarized towards science
(i.e., they belong in $\VlikesSc$), and we show the average of these
fractions for all of these users $u$. In the right panel, we do the
same, but considering the conspiracy users $u\in\VlikesCon$ and their
conspiracy friends.
Figure~\ref{fig:homophily} clearly indicates the presence of homophily
in the friendship network with respect to news preference.
This is an important phenomenon to understand viral processes, 
because the latter might diffuse among connected
users having similar information consumption patterns; we study this
question more in Section~\ref{sec:virality}. To further confirm such
point, we show that for a polarized user $u$, the fraction of polarized
friends in her category $y(u)$
can be predicted by means of a linear regression model with intercept
where the explanatory variable is a logarithmic transformation of the
number of likes $\NumLikes{u}$, that is, $y(u) = \beta_{0} +
\beta_{1}\log(\NumLikes{u}).$
Coefficients are estimated using ordinary
least squares and they are---with the corresponding standard
errors inside the round brackets---$\hat{\beta}_{0} = 0.70$ $(0.005)$ and
$\hat{\beta}_{1} = 0.043 $ $(0.001)$, with $R^{2}=0.95$, for users
polarized towards science, and $\hat{\beta}_{0} = 0.71$ $ (0.003)$ and
$\hat{\beta}_{1} = 0.047$ $ (0.0006)$, with $R^{2}=0.98$, for users
polarized towards conspiracy. 
All the p-values are close to zero.
In Figure \ref{fig:lr_homophily}, we show the fit of the model for users polarized in science and conspiracy.

Summarizing, we find very polarized communities around different
contents. Such a polarization emerges in the friendship network where we identify homophily;
indeed, the more a polarized user is active in her category the bigger
the number of polarized friends she has in the same category. Thanks to
such a characterization, we are able to identify areas of the social network
where a given content is more likely to diffuse. This is the topic of
the next section.

\section{Virality and Polarization}
\label{sec:virality}
In this section, we want to analyze the determinants behind viral
phenomena related to false information. In particular, we focus on the
role of structural features of the social network as well as on users'
consumption patterns.

\subsection{Network Structure and Users' Activity}

\begin{figure}
 \centering
	{\includegraphics[width=0.4\textwidth]{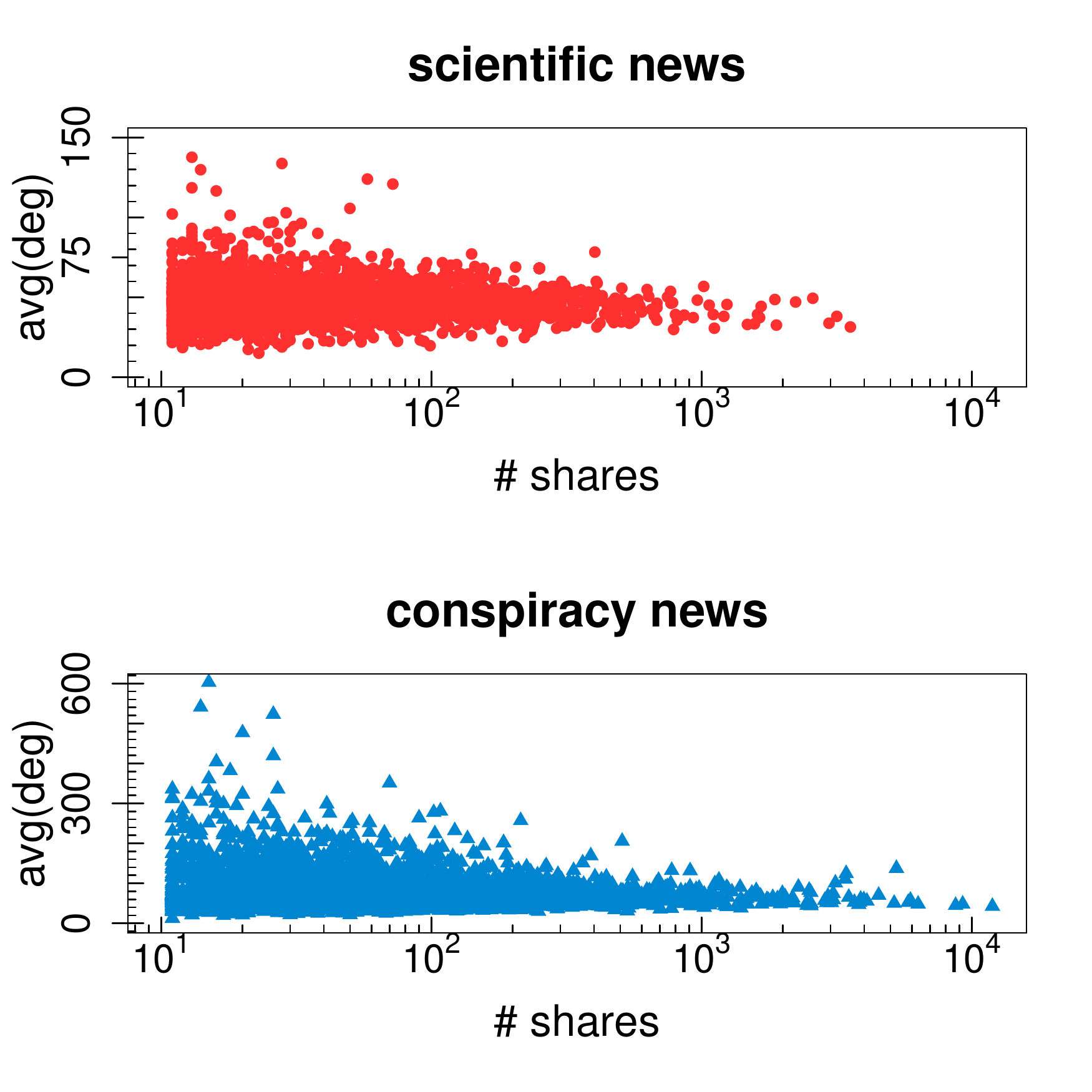}} 
\caption{Number of shares versus average degree of users who liked the posts in our samples.}
\label{fig:likers_shares_deg}
\end{figure}

\begin{figure}[H]
 \centering
	{\includegraphics[width=0.4\textwidth]{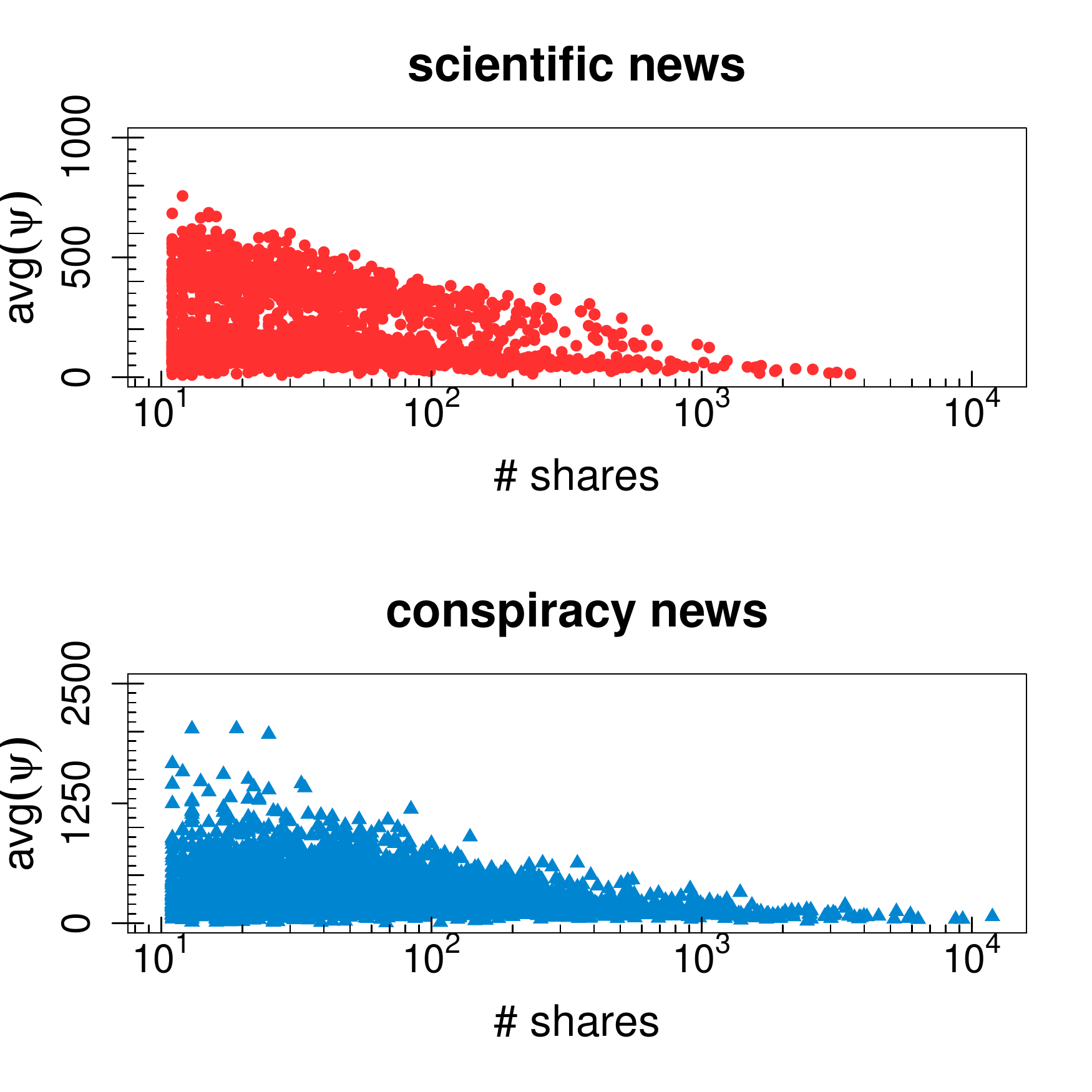}} 
\caption{Number of shares versus average liking activity of users who liked the posts in our samples.}
\label{fig:likers_shares_act}
\end{figure}

For a better understanding of posts' virality, we use the number of
shares since they are a proxy of the number of people reached by a post.
Hence, we pick two random samples of $5,000$ posts from each set $\PostsSc$ and
$\PostsCon$, and we compute structural as well as information-consumption
metrics to be compared with the total number of shares.  In
particular, we concentrate on the degree $\Degree{u}$ and liking activity
$\Engagement{u}$ of each user $u$ who liked the posts in our samples.  In
figures~\ref{fig:likers_shares_deg} and~\ref{fig:likers_shares_act}  we
show the number of shares versus,
respectively, the average values of the degree $\Degree{u}$ and liking
activity $\Engagement{u}$ among the users who liked a given post from our
samples. The results provide an outline of viral processes. They show
that in viral posts (i.e., posts with high values of shares), very active
users (in terms either of number of friends or liking activity) form a tiny fraction of the
overall users. We have checked that similar results hold when
accounting for commenting activity.
Since the distributions of both degree and liking activity are heavy
tailed, large volumes of users that make a post viral are mainly composed
of users with low degree and liking activity. 

\subsection{Viral Processes on False Information}

\begin{figure}
 \centering
\includegraphics[width=0.4\textwidth]{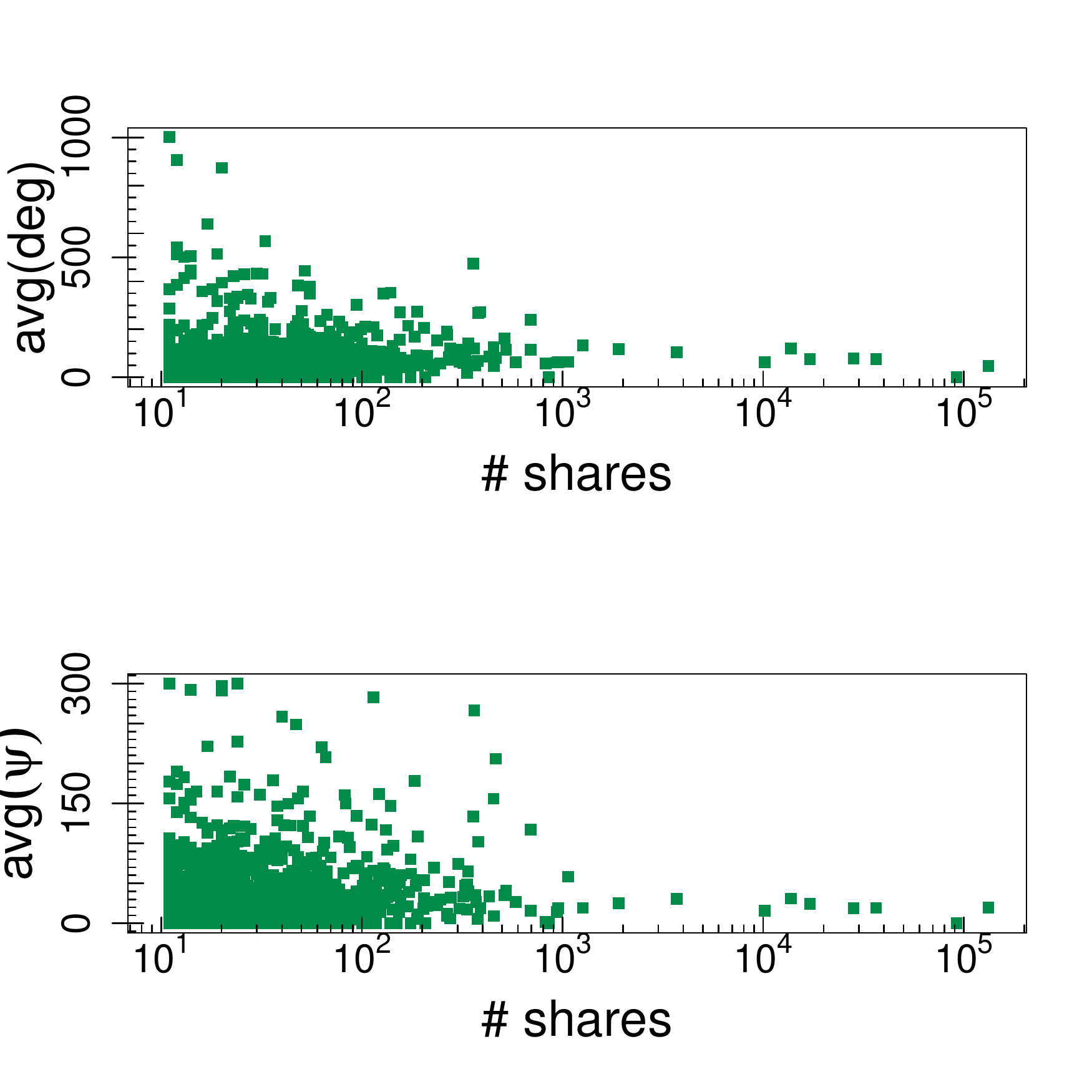}
\caption{Number of shares versus average degree and average liking
activity of users who liked troll posts.}
\label{fig:troll_virality}
\end{figure}

\begin{figure}
 \centering
       \includegraphics[width=0.4\textwidth]{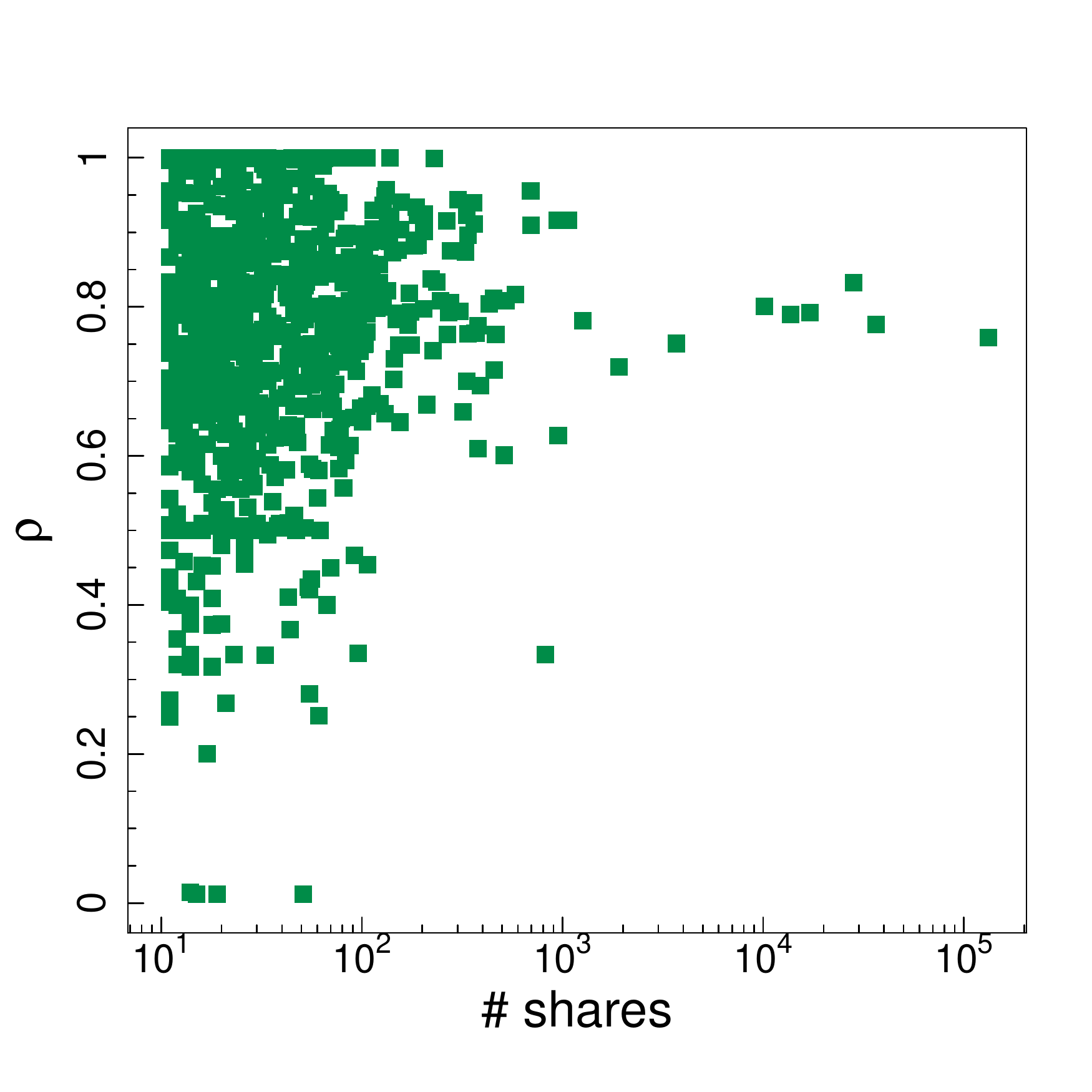}
\caption{Polarization on troll posts. The average polarization versus the number of shares for troll posts.}
\label{fig:troll_polarization}
\end{figure}

\begin{figure}
 \centering
       \includegraphics[width=0.4\textwidth]{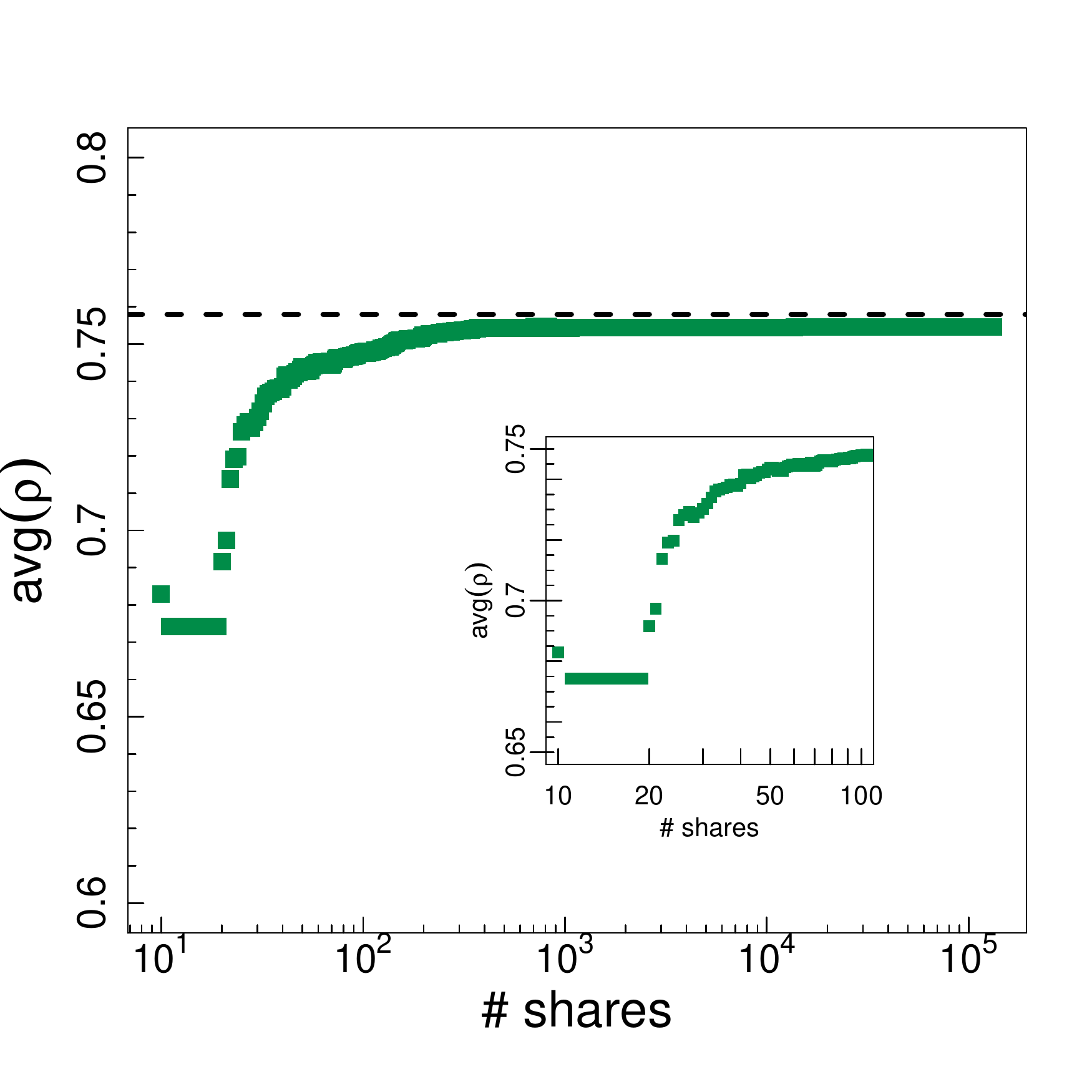}
\caption{Average polarization of users who liked troll posts for increasing values of the number of shares.}
\label{fig:troll_polarization2}
\end{figure}

Users' preference towards information belonging to a particular kind of
narrative might provide interesting insights with respect to the
determinants of viral phenomena.
Under the light of the findings in the previous section, we now focus on how
\emph{false information} can spread in \facebook.
We use the set of $4,709$ \emph{troll} posts described in
Section~\ref{sec:data} and we study the virality of each post with
respect to the network structure and users' liking activity. Notice that
such a set of posts is disjoint from the one used to classify users as
polarized towards science or conspiracy, thus we can obtain unbiased
results regarding the effect of users' polarization in the diffusion of
false information. Figure~\ref{fig:troll_virality} shows the same
information of figures~\ref{fig:likers_shares_deg} and~\ref{fig:likers_shares_act} but for troll posts.
Again, in viral posts, very active users (in terms either of number of friends or
liking activity) form a tiny fraction of the overall users.
However, we found an interesting fact: in most of the viral posts users
are highly polarized. In
Figure~\ref{fig:troll_polarization}, for each troll post, we show
polarization $\Polarization{\cdot}$ of users who liked it against its number of shares.
Whereas users' both degree and liking activity become low as the number
of shares increases, polarization $\Polarization{u}$ is
relevant for all levels of shares.
In Figure \ref{fig:troll_polarization2} we show the average value of
the polarization for increasing levels of the number of shares.
More precisely, for each level of virality $x$, we compute the average
polarization of all users who liked troll posts with number of shares at
least~$x$.
As we can see, the curve has an increasing trend that asymptotically
stabilizes around~$0.76$.
The plot indicates that for posts containing false claims
with more than 100 shares,
the average polarization of the users who liked
it is high and remains constant with the
increase (at different orders of magnitude) of the virality.

\begin{figure}
 \centering
       \includegraphics[width=0.4\textwidth]{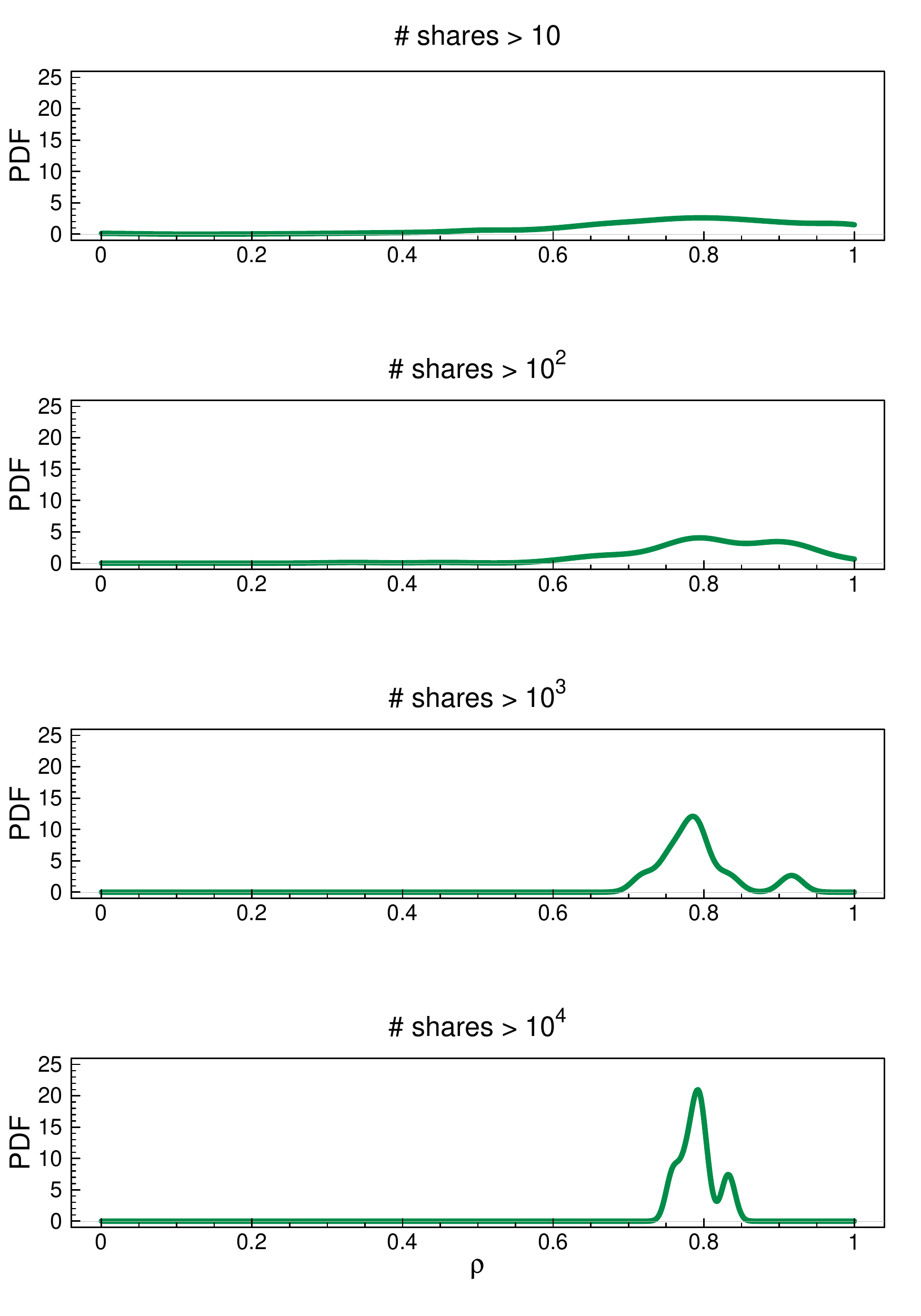}
\caption{PDF of the average polarization for different levels of virality.}
\label{fig:troll_polarization3}
\end{figure}

To better shape the role the polarization $\Polarization{\cdot}$ in
viral phenomena around false information, we study how it
changes by considering posts with increasing levels of virality.
In Figure~\ref{fig:troll_polarization3} we show the PDF of users'
average polarization for different levels of virality.
We observe that for highly viral posts (with false information) the
average polarization value of the users who like them peaks
around~$0.8$.

Observe that the pick increases with virality of posts.
indication that high polarization is most of the times essential for a
post becoming viral.
The results indicate that a good indicator to detect viral phenomena around false information is the users' polarization.
The diffusion of false claims proliferates within (tightly clustered) users that are usually exposed to unsubstantiated
rumors.

\subsection{Case Study}
%

\begin{figure}
 \centering
       \includegraphics[width=0.5\textwidth]{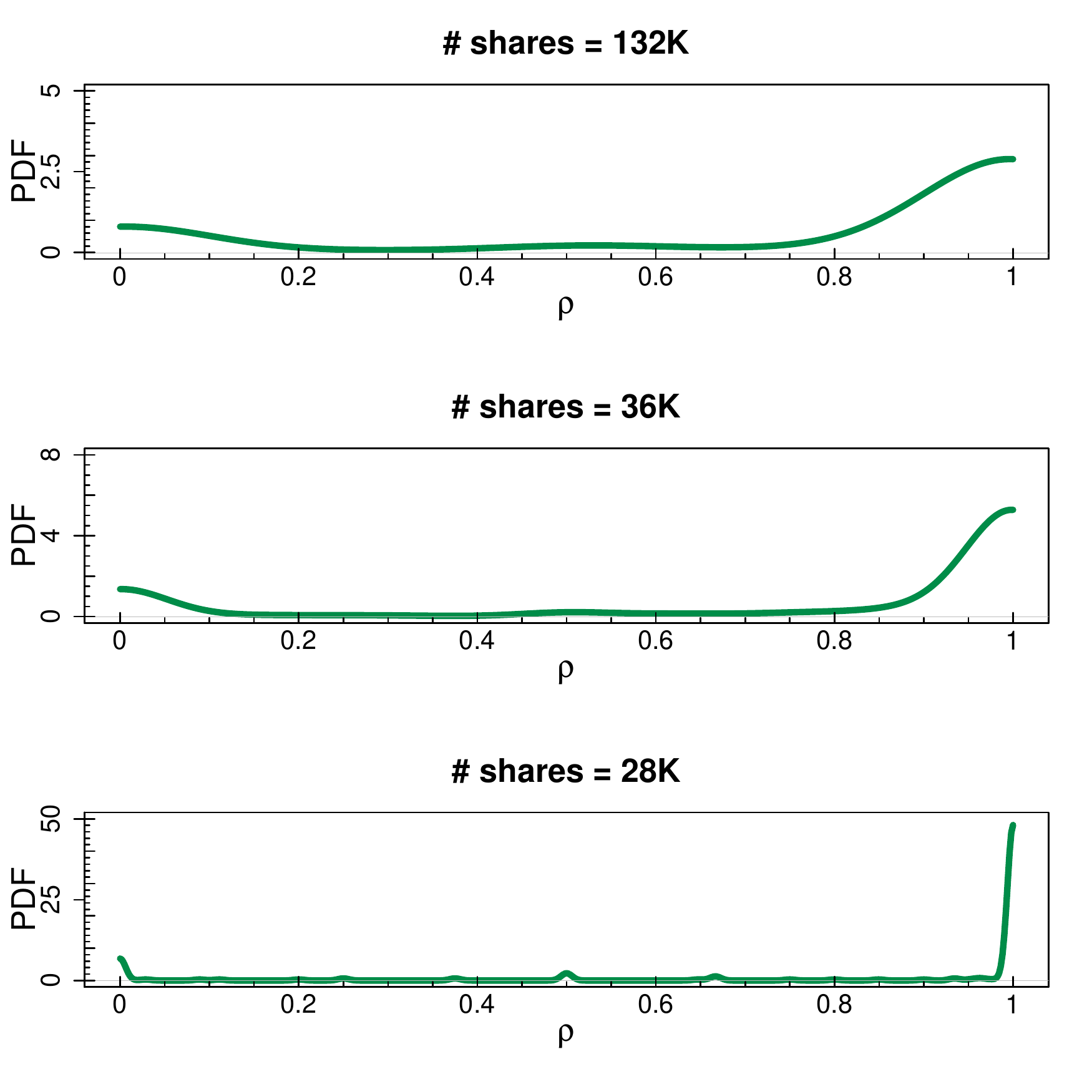}
\caption{PDF of the polarization of the users who liked one of the three
most viral troll posts.}
\label{fig:3_most_viral}
\end{figure}

Let us now obtain some more intuition by studying in more detail the
most viral troll posts~\cite{SM1,SM2,SM3}.
In Figure~\ref{fig:3_most_viral} we show the probability density function of the polarization
of the users that liked each of the three troll posts.
The most viral post with $132K$ shares says that in the year of the post
(2013), after $5,467$
years, the month of October has 5 Tuesdays, Wednesdays, and Thursdays, and
that this is a very rare event so that Chinese people call it year of
the glory \emph{shu tan tzu}. The fact that October 2013 has 5
Tuesdays, Wednesdays, and Thursdays is true, however the rest is false:
this happens around once every seven years and the phrase has no sense
in Chinese.
The second one is the popular case of Senator Cirenga mentioned in the
introduction, which received more than $36K$ shares.
The third one has a more political taste by stating that the former
Italian speaker of the house received a large sum of money after his
resignation. This post was shared $28K$ times.
In Figure~\ref{fig:3_most_viral} we show the PDF of users' polarization
liking these three posts.
We can see that in all three posts there is a significant proportion of
conspiracy users, who have seen the post through a viral process (note
that the total subscribers to the \facebook group pages are upper
bounded by about only $8K$, and we expect a much smaller proportion to
have seen the posts in their \facebook feeds). Interestingly, there is
always a no-negligible number of science users who have also liked the
posts, even though they are from four to ten times less.

\begin{figure}
 \centering
       \includegraphics[width=0.5\textwidth]{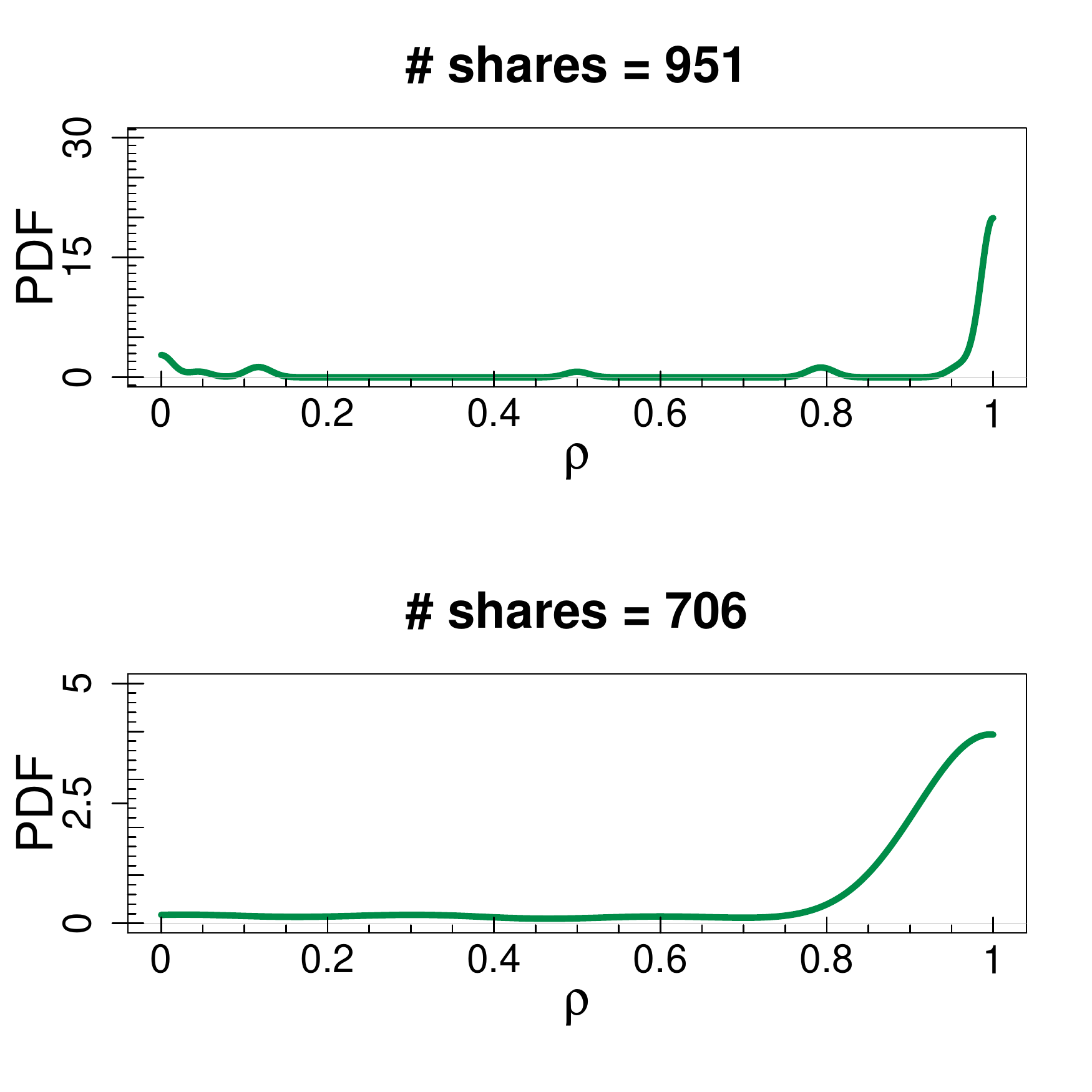}
\caption{PDF of the polarization of the users who liked one of the two
most viral conspiracy-like troll posts.}
\label{fig:most_viral_sh}
\end{figure}

Actually, we can observe that the percentage of science users depends
directly on the content of the post: the more towards conspiracy a post
is, the less the ratio of science users.
To provide a better outline of this phenomenon we also plot in Figure~\ref{fig:most_viral_sh}
the two most viral posts \cite{SH1,SH2} from the troll page that is more specific in
teasing conspiracy stories (see Section~\ref{sec:data}). The first post
mentions that a Chinese engineer discovered an eco-friendly (uranium-based) lamp that
produces extremely bright light at an extremely low cost (order of
pennies) while being able to several hundred years. The second one is a
version of the chemtrail conspiracy theory, mentioning, among others,
that one of the ingredients with which we are bing sprayed is {\em sildenafil
citrate}---the active ingredient of Viagra. Note that in those ones
the liking activity of science users is very close to zero.

\section{Conclusions}
Nowadays, understanding the determinants of viral phenomena occurring on
socio-technical systems is a very interesting challenge.
The free circulation of contents has changed the way people get
informed, and the quality of information plays a fundamental role in
collective debates.
In this work, we addressed the relationships between the virality of false
claims and the users information consumption patterns.
To do this, we analyze a sample of $1.2M$ Facebook Italian users
consuming different (and opposite) types of information---science and
conspiracy news. We found out that users engagement across
different contents correlates with the number of friends having similar
consumption patterns (\emph{homophily}). Next, we studied the
virality of posts of the two categories, finding that both present
similar statistical signatures. 
Finally, we tested diffusion patterns on an external set of $4,709$
intentionally satirical false posts (troll posts) and we discovered that
neither the presence of hubs (structural properties) nor the most active
users (influencers) are prevalent in viral phenomena. 
The results of this work provide important insights towards the
understanding of viral processes around false information on online
social media. 
In particular, we provide a new metric---user polarization defined on
the information consumption patterns---which along with 
users' engagement and the number of friends helps in the discovery of
where in the social network false claims are more likely to spread.
In addition, we show that users' aggregation around shared beliefs makes
the frequent exposure to conspiracy stories (\emph{polarization}) a determinant for the virality of 
false information.

\section{Acknowledgments}
Funding for this work was provided by EU FET project MULTIPLEX nr.
317532 and SIMPOL nr. 610704. The funders had no role in study design,
data collection and analysis, decision to publish, or preparation
of the manuscript.
Special thanks for building the atlas of pages to {\em Protesi di Protesi di Complotto}, {\em Che vuol dire reale}, {\em La menzogna diventa verita e passa alla storia}, {\em Simply Humans}, {\em Semplicemente me}.
Finally, we want to mention Salvatore Previti, Elio Gabalo, Titta, Morocho, and Sandro Forgione for interesting discussions.

%

\balancecolumns
\end{document}